\documentclass[a4paper,11pt]{article}
\usepackage[utf8]{inputenc}
\pdfoutput=1
\usepackage{jcappub}

\usepackage{xcolor}
\usepackage{amssymb}
\usepackage{amsmath}
\usepackage{tabularx,booktabs} 
\usepackage{multirow}
\usepackage{todonotes}

\title{Results of the follow-up of ANTARES neutrino alerts}

\author[1,2]{A.~Albert}
\author[3]{S.~Alves}
\author[4]{M.~Andr\'e}
\author[5]{M.~Ardid}
\author[5]{S.~Ardid}
\author[6]{J.-J.~Aubert}
\author[7]{J.~Aublin}
\author[7]{B.~Baret}
\author[8]{S.~Basa}
\author[7]{Y.~Becherini}
% ...

\author{(ANTARES Collaboration\footnote[1]{The shown author affiliations reflect their job contracts; the ANTARES collaboration has currently suspended all institutional relations with Russian science organisations.})\\}
\author[43]{\\ V. Lipunov}
\author[43]{G.Antipov}
\author[43]{P.Balanutsa}
\author[45,46,47,48]{D.Buckley}
\author[44]{N.Budnev}
\author[43]{A.Chasovnikov}
\author[43]{D. Cheryasov}
\author[50,51]{C.Francile}
\author[43]{A.Gabovich}
\author[43]{E.Gorbovskoy}
\author[43]{I. Gorbunov}
\author[43]{O. Gress}
\author[43]{V. Kornilov}
\author[43]{A. Kuznetsov}
\author[43]{A. Iyudin}
\author[50,51]{R.Podesta}
\author[50,51]{F.Podesta}
\author[49]{R.Rebolo Lopez}
\author[43]{V.Senik}
\author[49]{M.Sierra-Rucart}
\author[43]{S. Svertilov}
\author[43]{N.Tiurina}
\author[43]{D. Vlasenko}
\author[43]{I. Yashin}
\author[43]{K. Zhirkov}
\author{(Master Collaboration)\\}
\author[52,53]{\\ S. Croft}
\author[54]{D.L. Kaplan}
\author[55]{G.E. Anderson}
\author[55]{A. Williams}
\author[56,57]{D. Dobie}
\author[58]{K. W. Bannister}
\author[59,60]{P.~J.~Hancock}
\author{(M.W.A. Collaboration)\\}
\author[61]{\\ P.A. Evans} 
\author[62]{J.A. Kennea}
\author[61]{J.P. Osborne}
\author[63]{S.B. Cenko}
\author{(Swift Collaboration)\\}
\author[65]{\\ S. Antier}
\author[64]{J.L. Atteia}
\author[65]{M. Bo\"er}
\author[64,66]{A. Klotz}
\author{(TAROT Collaboration)\\}
\author[7]{\\ S. Chaty}
\author[67]{Klaus Hodapp}
\author[68,69]{\\V. Savchenko}

\affiliation[1]{\scriptsize{Universit\'e de Strasbourg, CNRS,  IPHC UMR 7178, F-67000 Strasbourg, France}}
\affiliation[2]{\scriptsize Universit\'e de Haute Alsace, F-68100 Mulhouse, France}
\affiliation[3]{\scriptsize{IFIC - Instituto de F\'isica Corpuscular (CSIC - Universitat de Val\`encia) c/ Catedr\'atico Jos\'e Beltr\'an, 2 E-46980 Paterna, Valencia, Spain}}
\affiliation[4]{\scriptsize{Technical University of Catalonia, Laboratory of Applied Bioacoustics, Rambla Exposici\'o, 08800 Vilanova i la Geltr\'u, Barcelona, Spain}}
\affiliation[5]{\scriptsize{Institut d'Investigaci\'o per a la Gesti\'o Integrada de les Zones Costaneres (IGIC) - Universitat Polit\`ecnica de Val\`encia. C/  Paranimf 1, 46730 Gandia, Spain}}
\affiliation[6]{\scriptsize{Aix Marseille Univ, CNRS/IN2P3, CPPM, Marseille, France}}
\affiliation[7]{\scriptsize{Universit\'e Paris Cit\'e, CNRS, Astroparticule et Cosmologie, F-75013 Paris, France}}
\affiliation[8]{\scriptsize{Aix Marseille Univ, CNRS, CNES, LAM, Marseille, France }}
\affiliation[9]{\scriptsize{National Center for Energy Sciences and Nuclear Techniques, B.P.1382, R. P.10001 Rabat, Morocco}}
\affiliation[10]{\scriptsize{University Mohammed V in Rabat, Faculty of Sciences, 4 av. Ibn Battouta, B.P. 1014, R.P. 10000
Rabat, Morocco}}
\affiliation[11]{\scriptsize{INFN - Sezione di Bologna, Viale Berti-Pichat 6/2, 40127 Bologna, Italy}}
\affiliation[12]{\scriptsize{Dipartimento di Fisica e Astronomia dell’Università di Bologna, Viale Berti-Pichat 6/2, 40127, Bologna, Italy}}
\affiliation[13]{\scriptsize{INFN - Laboratori Nazionali del Sud (LNS), Via S. Sofia 62, 95123 Catania, Italy}}
\affiliation[14]{\scriptsize{Friedrich-Alexander-Universit\"at Erlangen-N\"urnberg, Erlangen Centre for Astroparticle Physics, Erwin-Rommel-Str. 1, 91058 Erlangen, Germany}}
\affiliation[15]{\scriptsize{University Mohammed I, Laboratory of Physics of Matter and Radiations, B.P.717, Oujda 6000, Morocco}}
\affiliation[16]{\scriptsize{Nikhef, Science Park,  Amsterdam, The Netherlands}}
\affiliation[17]{\scriptsize{Institute of Space Science, RO-077125 Bucharest, M\u{a}gurele, Romania}}
\affiliation[18]{\scriptsize{Universiteit van Amsterdam, Instituut voor Hoge-Energie Fysica, Science Park 105, 1098 XG Amsterdam, The Netherlands}}
\affiliation[19]{\scriptsize{INFN - Sezione di Genova, Via Dodecaneso 33, 16146 Genova, Italy}}
\affiliation[20]{\scriptsize{INFN - Sezione di Roma, P.le Aldo Moro 2, 00185 Roma, Italy}}
\affiliation[21]{\scriptsize{Dipartimento di Fisica dell'Universit\`a La Sapienza, P.le Aldo Moro 2, 00185 Roma, Italy}}
\affiliation[22]{\scriptsize{LPHEA, Faculty of Science - Semlali, Cadi Ayyad University, P.O.B. 2390, Marrakech, Morocco.}}
\affiliation[23]{\scriptsize{INFN - Sezione di Bari, Via E. Orabona 4, 70126 Bari, Italy}}
\affiliation[24]{\scriptsize{International Centre for Radio Astronomy Research, Curtin University, Bentley, WA 6102, Australia}}
\affiliation[25]{\scriptsize{Department of Computer Architecture and Technology/CITIC, University of Granada, 18071 Granada, Spain}}
\affiliation[26]{\scriptsize{Universit\'e Paris-Sud, 91405 Orsay Cedex, France}}
\affiliation[27]{\scriptsize{Universit\`a di Salerno e INFN Gruppo Collegato di Salerno, Dipartimento di Fisica, Via Giovanni Paolo II 132, Fisciano, 84084 Italy}}
\affiliation[28]{\scriptsize{Laboratoire de Physique Corpusculaire, Clermont Universit\'e, Universit\'e Blaise Pascal, CNRS/IN2P3, BP 10448, F-63000 Clermont-Ferrand, France}}
\affiliation[29]{\scriptsize{LIS, UMR Universit\'e de Toulon, Aix Marseille Universit\'e, CNRS, 83041 Toulon, France}}
\affiliation[30]{\scriptsize{Dipartimento di Fisica dell'Universit\`a, Via Dodecaneso 33, 16146 Genova, Italy}}
\affiliation[31]{\scriptsize{Royal Netherlands Institute for Sea Research (NIOZ), Landsdiep 4, 1797 SZ 't Horntje (Texel), the Netherlands}}
\affiliation[32]{\scriptsize{G\'eoazur, UCA, CNRS, IRD, Observatoire de la C\^ote d'Azur, Sophia Antipolis, France}}
\affiliation[33]{\scriptsize{Huygens-Kamerlingh Onnes Laboratorium, Universiteit Leiden, The Netherlands}}
\affiliation[34]{\scriptsize{Institut f\"ur Theoretische Physik und Astrophysik, Universit\"at W\"urzburg, Emil-Fischer Str. 31, 97074 W\"urzburg, Germany}}
\affiliation[35]{\scriptsize{Dr. Remeis-Sternwarte and ECAP, Friedrich-Alexander-Universit\"at Erlangen-N\"urnberg,  Sternwartstr. 7, 96049 Bamberg, Germany}}
\affiliation[36]{\scriptsize{Mediterranean Institute of Oceanography (MIO), Aix-Marseille University, 13288, Marseille, Cedex 9, France; Universit\'e du Sud Toulon-Var,  CNRS-INSU/IRD UM 110, 83957, La Garde Cedex, France}}
\affiliation[37]{\scriptsize{INFN - Sezione di Catania, Via S. Sofia 64, 95123 Catania, Italy}}
\affiliation[38]{\scriptsize{IRFU, CEA, Universit\'e Paris-Saclay, F-91191 Gif-sur-Yvette, France}}
\affiliation[39]{\scriptsize{INFN - Sezione di Napoli, Via Cintia 80126 Napoli, Italy}}
\affiliation[40]{\scriptsize{Dipartimento di Fisica dell'Universit\`a Federico II di Napoli, Via Cintia 80126, Napoli, Italy}}
\affiliation[41]{\scriptsize{Dpto. de F\'\i{}sica Te\'orica y del Cosmos \& C.A.F.P.E., University of Granada, 18071 Granada, Spain}}
\affiliation[42]{\scriptsize{Institut Universitaire de France, 75005 Paris, France}}

\affiliation[43]{\scriptsize{M.V.Lomonosov Moscow State University, 119234, 1, Leninskie Gory, Moscow, Russia}}
\affiliation[44]{\scriptsize{Irkutsk State University, Applied Physics Institute, 20, Gagarin blvd,664003, Irkutsk, Russia}}
\affiliation[45]{\scriptsize{South African Astronomical Observatory, PO Box 9, Observatory 7935, Cape Town, South Africa}}
\affiliation[46]{\scriptsize{Southern African Large Telescope, P.O. Box 9, Observatory, 7935, Cape Town, South Africa}}
\affiliation[47]{\scriptsize{Department of Astronomy, University of Cape Town, Private Bag X3, Rondebosch 7701, South Africa}}
\affiliation[48]{\scriptsize{Department of Physics, University of the Free State, PO Box 339, Bloemfontein 9300, South Africa}}
\affiliation[49]{\scriptsize{Instituto de Astrofisica de Canarias Via Lactea, s/n E38205 - La Laguna (Tenerife), Spain}}
\affiliation[50]{\scriptsize{San Juan National University, Casilla de Correo 49, San Juan 5400, Argentina}}
\affiliation[51]{\scriptsize{Observatorio Astronomico Felix Aguilar (OAFA), Avda Benavides 8175, Rivadavia, El Leonsito, San Juan 5400, Argentina}}

\affiliation[52]{\scriptsize{Department of Astronomy, University of California, Berkeley, 501 Campbell Hall 3411, Berkeley, CA, 94720, USA}}
\affiliation[53]{\scriptsize{SETI Institute, 339 N Bernardo Ave, Mountain View, CA, 94043, USA}}
\affiliation[54]{\scriptsize{Center for Gravitation, Cosmology, and Astrophysics, Department of Physics, University of Wisconsin-Milwaukee, P.O. Box 413, Milwaukee, WI 53201, USA}}
\affiliation[55]{\scriptsize{International Centre for Radio Astronomy Research, Curtin University, GPO Box U1987, Perth, WA 6845, Australia}}
\affiliation[56]{\scriptsize{Centre for Astrophysics and Supercomputing, Swinburne University of Technology, Hawthorn, Victoria, Australia}}
\affiliation[57]{\scriptsize{ARC Centre of Excellence for Gravitational Wave Discovery (OzGrav), Hawthorn, Victoria, Australia}}
\affiliation[58]{\scriptsize{CSIRO, Space and Astronomy, PO Box 76, Epping, NSW 1710, Australia}}
\affiliation[60]{\scriptsize{Curtin Institute for Computation, Curtin University, GPO Box U1987, Perth WA 6845}}

\affiliation[61]{\scriptsize{School of Physics \& Astronomy, University of Leicester, University Road, Leicester, LE1 7RH, UK}}
\affiliation[62]{\scriptsize{Department of Astronomy and Astrophysics, The Pennsylvania State University, 525 Davey Lab, University Park, PA 16802, USA}}
\affiliation[63]{\scriptsize{Astrophysics Science Division, NASA Goddard Space Flight Center, 8800 Greenbelt Rd, Greenbelt, MD 20771, USA   and Joint Space-Science Institute, University of Maryland, College Park, Maryland 20742, USA}}

\affiliation[64]{\scriptsize{IRAP, Universit\'e de Toulouse, CNRS, CNES, UPS, (Toulouse), France}}
\affiliation[65]{\scriptsize{Universit\'e Co\^te d’Azur, Observatoire de la Co\^te d’Azur, CNRS, Artemis, Boulevard de l’Observatoire, 06304 Nice, France}}
\affiliation[66]{\scriptsize{Universit\'e de Toulouse; UPS-OMP; IRAP; 31400 Toulouse, France}}

\affiliation[67]{\scriptsize{Polish Academy of Sciences, Nicolaus Copernicus Astronomical Center, Bartycka 18, 00-716 Warszawa, Poland;
Ruhr University Bochum, Faculty of Physics and Astronomy, Astronomical Institute (AIRUB), 44780 Bochum, Germany;
Universidad Cat\'olica del Norte, Instituto de Astronomía, Avenida Angamos 0610, Antofagasta, Chile}}

\affiliation[68]{\scriptsize{EPFL Laboratoire d’astrophysique, Observatoire de Sauverny, CH-1290 Versoix, Switzerland}}
\affiliation[69]{\scriptsize{D\'partement d'Astronomie Universit\' de Gen\`ve, Observatoire de Sauverny, CH-1290 Versoix, Switzerland}}

\emailAdd{dornic@cppm.in2p3.fr; coleiro@apc.in2p3.fr}

\abstract{
High-energy neutrinos could be produced in the interaction of charged cosmic rays with matter or radiation surrounding astrophysical sources. To look for transient sources associated with neutrino emission, a follow-up program of neutrino alerts has been operating within the ANTARES Collaboration since 2009. This program, named TAToO, has triggered robotic optical telescopes (MASTER, TAROT, ROTSE and the SVOM ground based telescopes) immediately after the detection of any relevant neutrino candidate and scheduled several observations in the weeks following the detection. A subset of ANTARES events with highest probabilities of being of cosmic origin has also been followed by the Swift and the INTEGRAL satellites, the Murchison Widefield Array radio telescope and the H.E.S.S. high-energy gamma-ray telescope. The results of twelve years of observations are reported. No optical counterpart has been significantly associated with an ANTARES candidate neutrino signal during image analysis. Constraints on transient neutrino emission have been set. In September 2015, ANTARES issued a neutrino alert and during the follow-up, a potential transient counterpart was identified by Swift and MASTER. A multi-wavelength follow-up campaign has allowed to identify the nature of this source and has proven its fortuitous association with the neutrino. The return of experience is particularly important for the design of the alert system of KM3NeT, the next generation neutrino telescope in the Mediterranean Sea.}

\begin{document}
\maketitle
\flushbottom

\section{Introduction}
Neutrino astronomy allows the study of some of the most energetic non-thermal sources in the Universe. Despite the observations of cosmic rays up to ultra-high energies, of $\gamma$-rays and of a diffuse astrophysical neutrino flux, where or how these particles are accelerated is still unknown. The observation of neutrinos from peculiar directions, or in coincidence with transient phenomena, is clearly a key to directly answer these questions. Astrophysical neutrinos provide insight into source characteristics not accessible through the observation of other messengers. Due to their low cross section, neutrinos can escape dense astrophysical environments that are opaque to photons. In contrast to $\gamma$-rays, neutrinos travel through the Universe almost without interactions, allowing direct observation of their sources at high redshift with sub-degree-scale pointing precision. Unlike charged cosmic rays, neutrinos are not deflected by magnetic fields and can be observed in spatial and temporal coincidences with photons and gravitational waves, which is a key prerequisite to reap the scientific rewards of multi-messenger astronomy.\\

Doing astronomy with neutrinos is a long-standing aspiration. Up to now, IceCube at the South Pole~\cite{Aartsen:2016nxy}, ANTARES in the Mediterranean Sea~\cite{Collaboration:2011nsa} and GVD in the Lake Baikal~\cite{2020PAN....83..916A} have been the main players for neutrino astronomy analyses. The last decade was marked by the IceCube results in high-energy neutrino astronomy, with the discovery of an astrophysical diffuse neutrino flux in the 10 TeV – 10 PeV energy range using high-energy starting events (HESE)~\cite{Aartsen:2014gkd}. On September 22, 2017, the IceCube Collaboration emitted a public alert for the HESE event IC-170922A. Following the alert, Fermi-LAT and MAGIC detected an increased gamma-ray flux from the known blazar TXS~0506+056~\cite{IceCube:2018dnn,IceCube:2018cha} compatible with the direction of the IceCube neutrino alert. The probability of such an association reaches a 3$\sigma$ level. The TXS 0506+056 direction was also investigated by ANTARES in steady and time-dependant modes with 11 years of data~\cite{Albert:2018kjg}. No excess of events over the expected background was found in these two analyses. \\

In this context, multi-messenger approaches consisting of simultaneously looking for the same sources with neutrino telescopes, gravitational wave interferometers and/or multi-wavelength facilities constitute a privileged way of identifying astrophysical cosmic-ray accelerators. An alert system, dubbed TAToO (Telescopes-Antares Target of Opportunity), has been operating since 2009~\cite{Ageron:2011pe}, sending alerts to partners operating classical (electromagnetic, EM) telescopes. This approach does not require an \textit{a priori} hypothesis on the nature of the underlying neutrino source. It relies only on the hypothesis that these astrophysical phenomena produce high-energy neutrino and electromagnetic radiation over a broad energy range. In particular, the system mainly targets very fast transient sources such as gamma-ray bursts (GRBs) or quite long-term variable sources such as core-collapse supernovae (CCSN), and the flares of active galactic nuclei (AGN). \\

The rapid provision of alerts for interesting neutrino events enables both ground- and space-based observatories to quickly point at the direction of the alert. This fast follow-up is vital to catch and characterise any multi-messenger and multi-wavelength counterparts of these cataclysmic and short-lived phenomena. By combining the information provided by the ANTARES neutrino telescope with information coming from other observatories, the probability of detecting a source is enhanced, allowing the possibility of identifying a neutrino progenitor from a single detected event. The gain in sensitivity can be as large as a factor of ten compared to a steady point-like source analysis. In this respect, this program offers a mutual benefit to all partners. The first results of the early follow-up observations have been published in Ref.~\cite{Adrian-Martinez:2015nin}.\\

\begin{figure}[!ht]
    \centering
    \includegraphics[width=0.9\linewidth]{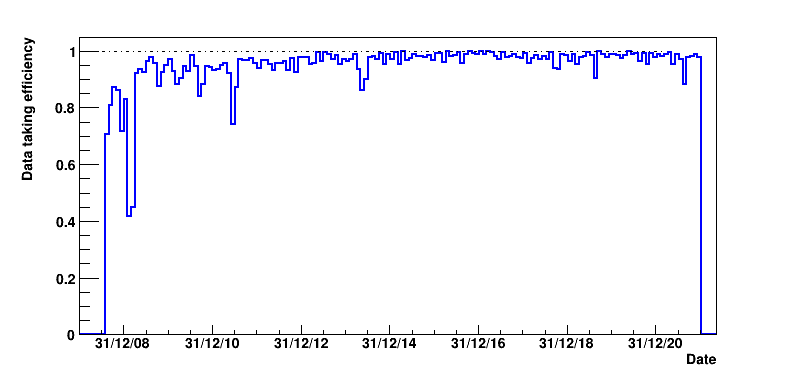}
    \caption{Fraction of the monthly livetime of ANTARES online data-taking between 2008 and 2022.}
    \label{fig:eff}
\end{figure}

Completed in 2008, ANTARES (Astronomy with a Neutrino Telescope and Abyss Environmental Research) is the first neutrino telescope installed in the Mediterranean Sea~\cite{Collaboration:2011nsa}. The data acquisition was definitively stopped on February 12$^\mathrm{th}$, 2022 and the detector decommissioning started in May 2022. The detector was composed of 12 detection lines of about 500~m height anchored at a depth of 2500~m offshore Toulon (42$^\circ$48$^\prime$N, 6$^\circ$10$^\prime$E). The mean distance between lines was about 65 m. Each line was formed by a chain of 25 storeys with an inter-storey distance of 14.5 m. Every storey holded three optical modules housing a single 10-inch photomultiplier tube (PMT) looking downward at an angle of 45$^{\circ}$. In total, a mass of about 10 Mton of water was instrumented with 885 optical modules. The fraction of the monthly livetime of the detector was on average larger than 94\% (Figure~\ref{fig:eff}). The data losses were due to the shutdown of data taking, calibration periods or too high bioluminescence activity.\\ 


 The paper describes the main results of the follow-ups of neutrino ANTARES alerts since 2009. Section 2 summarises the alert system while Section 3 introduces the EM partners. Section 4 details the main results of this program. Conclusions and outlooks are drawn in Section 5.

\section{Alert system}

In parallel to the ANTARES acquisition chain, a fast and robust online algorithm reconstructed all incoming events in nearly real time \cite{Aguilar:2011zz}. A pre-selection was made based on the results of this fast reconstruction (upgoing events and quality cuts) to reduce the rate of events (from $\sim3$ Hz to $\sim0.01$ Hz). The selected sample of events was then sent to a more accurate reconstruction algorithm \cite{AdrianMartinez:2012rp}, which improved the angular precision of each event. This additional step delayed the event processing by a few seconds. In the online framework, both algorithms used an idealised geometry of the detector that did not take into account the dynamical positioning of the optical modules. The monitoring of the variation of the position of the optical modules was necessary due to the presence of variable underwater sea currents~\cite{2012JInst...7T8002A}. For high-energy tracks, online reconstruction algorithms reached a median angular resolution of $\sim$~0.5$^\circ$. However, in the case of sea-current speeds above 10~cm/s, the line deformations were such that the reconstruction algorithms did not perform accurately and therefore the reconstructed direction could not be used anymore for the alert sending. This situation happened typically on a $\sim$~15\% fraction of the ANTARES duty cycle (in addition to the data taking efficiency presented in Figure~\ref{fig:eff}).\\

From the previously described selected neutrino sample, the neutrino candidates with an increased probability to be of cosmic origin were singled out using the zenith direction (only upgoing events), the track reconstruction quality as well as the number of PMT hits selected by the triggers and the total amplitude of the PMT hits further selected by the reconstruction. A hit is a PMT signal with an amplitude above 0.3 photoelectrons. These two last parameters were used as an energy proxy. Four online neutrino trigger criteria have been implemented in the TAToO alert system~\cite{Ageron:2011pe, Adrian-Martinez:2015nin}:
\begin{itemize}
\item{High energy (HE) trigger: the detection of a single high-energy neutrino with a typical energy $\geq$ 5 TeV. The rate was about 1 per month.}
\item{Very high energy (VHE) trigger: the detection of a single very high-energy neutrino with an energy $\geq$ 30 TeV. This sub-sample of the HE trigger had a typical rate of 3$-$5 events per year.}
\item{Directional: the detection of a single neutrino for which the direction points toward ($\leq$ 0.4$^{\circ}$) a local galaxy ($\leq$ 20 Mpc) in the Gravitational Wave Galaxy Catalogue~\cite{White:2011qf}. This trigger was mainly introduced to enhance the chance to detect a local CCSN. The typical rate was about one per month.}
\item{Doublet trigger: the detection of at least two neutrinos coming from close directions ($\leq$ 3$^{\circ}$) within a predefined time window (15 min). No doublet trigger has ever been issued.}
\end{itemize}

 The trigger conditions were inspired by the features expected from astrophysical sources and were tuned to comply with the alert rate requested by the telescope networks. 
 Based on the distribution of the number of hits and of  the total amplitude of these hits for the different trigger types, a p-value corresponding to the fraction of online events with an energy equal or above the measured one in the full ANTARES dataset, is built. An agreement between ANTARES and the optical telescope collaborations allowed a rate of around up to 25 alerts per year to be sent, while an agreement to send up to 6 alerts per year to the Swift satellite was accepted. \\

\begin{figure}[!ht]
\centering
  \includegraphics[width=0.6\linewidth]{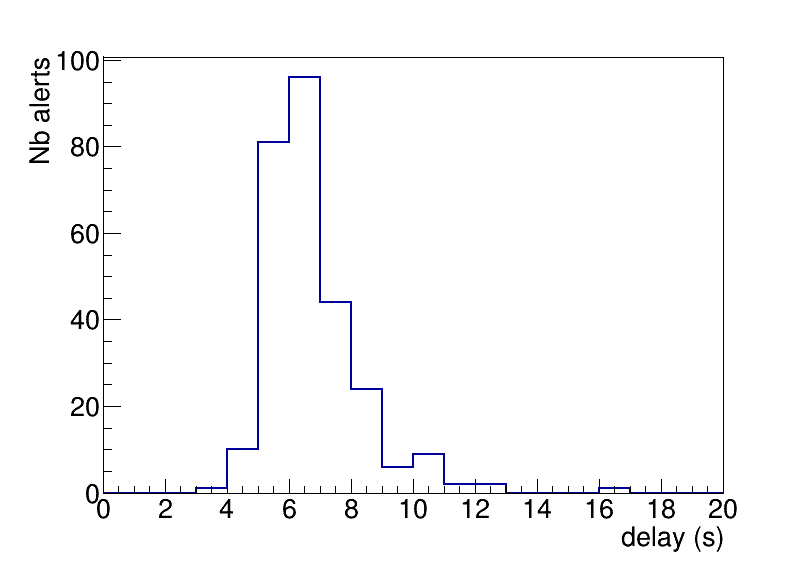}  
  \caption{Distribution of delays for alerts in the period 2011 and 2022 between the time of the neutrino detection and the corresponding alert transmission.}
  \label{fig:perfotime}
\end{figure}

\begin{figure}[!ht]
\centering
  \includegraphics[width=0.45\linewidth]{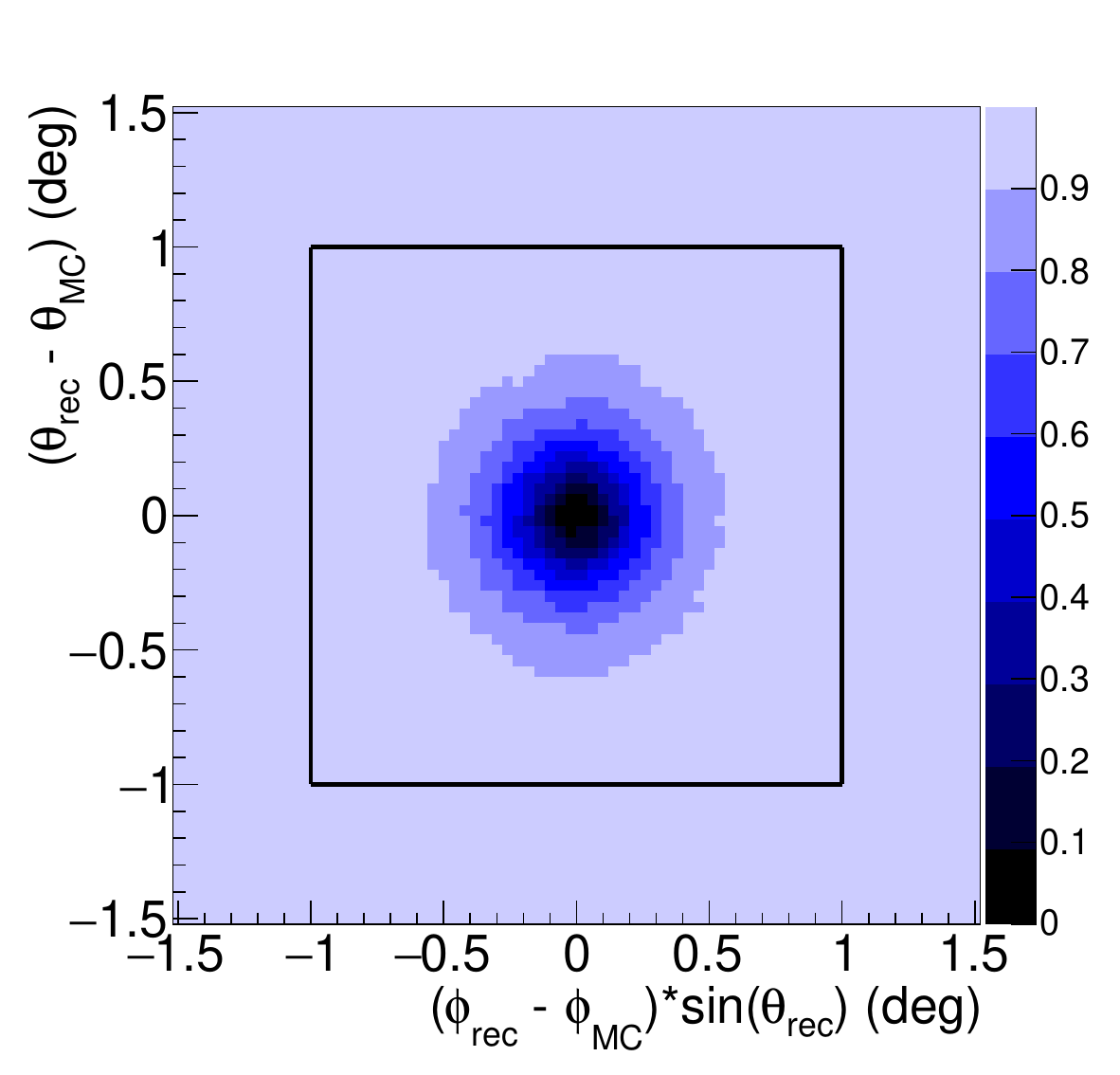} 
  \includegraphics[width=0.45\linewidth]{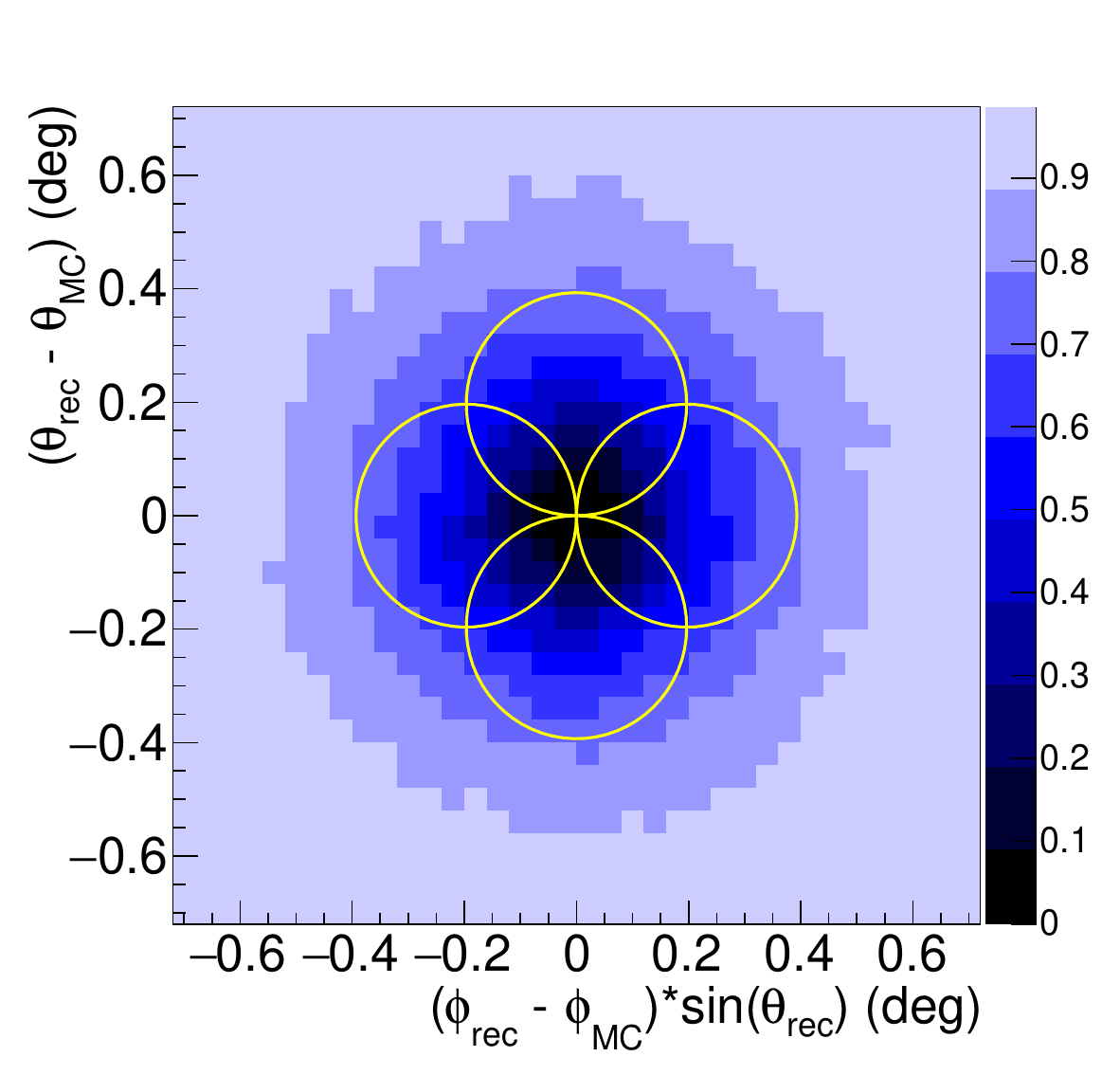}
  \caption{(left) Bi-dimensional angular resolution for a typical high-energy neutrino alert. (right) Zoom of the centre. The vertical scale is normalised to 1. The black square and circles correspond to the TAROT/ROTSE telescope and Swift/XRT fields of view, respectively. The 4 circles correspond to the tiling strategy used for the Swift/XRT follow-up (See Section~3). ($\theta_{\mathrm{rec}}$, $\phi_{\mathrm{rec}}$) and ($\theta_{\mathrm{MC}}$, $\phi_{\mathrm{MC}}$) are the reconstructed and the true MC local coordinates of the neutrino events. }
  \label{fig:psf}
\end{figure}

The TAToO alert system was able to send alerts within a few seconds after the neutrino detection with a localisation accuracy of about 0.5$^{\circ}$ (radius, 50\% containment). Figure~\ref{fig:perfotime} displays the distribution of the delays between the time at which the neutrinos are detected and the time of the associated alert message. This delay accounts for the time to collect all the hits to the shore station, the filtering and the fast reconstruction, the second, more accurate, reconstruction and finally the processing of the alert message. Figure~\ref{fig:psf} shows the Monte Carlo estimate of the point spread function for a typical HE neutrino alert, compared to the fields of view (FoV) of TAROT/ROTSE and Swift/XRT telescopes. \\


The online event processing and the alert distribution were managed by a custom application~\cite{Ageron:2011pe}, which periodically checked the connections to the networks and, in case of failure, automatic reconnection was performed, resulting in a fully autonomous and stable system. A veto prevented an alert to be sent if the ANTARES event counting rate exceeded a given threshold. In addition, if the alert criteria were fulfilled soon after a previous alert had already been issued, the new alert was stored in a FIFO and sent only after a certain period of time. This time lag was set to one hour, to avoid alert pileup in the optical telescope network. All alerts were sent using the Gamma-ray bursts Coordinates Network (GCN)~\cite{GCN2000} normalised format and using the standard VO Event format (XML file~\cite{2011ivoa.spec.0711S}). Information about the event that triggered the alert, i.e., a unique identifier, the time and the celestial coordinates, and the event p-value were sent to our partners at the time of the alert. The alerts were named ANTyymmddA using the same convention as for GRBs (in the case of the first alert of the day).\\


\section{Partner followers}
The follow-up of the ANTARES alerts started with the robotic optical telescopes TAROT and ROTSE in 2009. Their wide fields of view and their fast responses (images can be taken within less than 20 s after the neutrino alert) are well suited to search for transient sources. TAROT~\cite{Klotz_2009} is a network of two identical 0.25 m telescopes with a FoV of about 1.86$^\circ\times$1.86$^\circ$, located in Calern (France) and La Silla (Chile). The typical exposure is 60 seconds with a clear filter. Until the end of 2014, the network also comprised the four optical telescopes ROTSE~\cite{ROTSE2003}, which have progressively stopped their activity. ROTSE telescopes had similar properties as the TAROT ones. Since 2015, the MASTER network~\cite{2003AAS...202.4702L, 2005Ap.....48..389L} has also been observing the ANTARES neutrino alerts. It is composed of 8 observatories located in Russia (MASTER-Amur, -Tunka, -Ural, -Kislovodsk, -Tavrida), Canary Islands (-IAC), Argentina (-OAFA) and South Africa (-SAAO). Each observatory contains a twin 0.4 m telescope hosting one wide field (4 or 8 deg$^{2}$) and one very wide field (2$\times$400 deg$^{2}$) optical channels installed on a fast mount (up to 30$^\circ$ per sec) \cite{2010AdAst2010E..30L, 2012ExA....33..173K, 2020ApJ...896L..19L}. The typical exposure time of the images is 60 or 180 seconds, depending on the current moon phase. MASTER automatically follows the alerts and analyses the images. An auto-detection system allows for a fast transmission of the optical transients~\cite{2010AdAst2010E..30L,2020ApJ...896L..19L}. These telescopes reach a limiting magnitude of about 19$-$20.5 mag depending on the observing conditions. Zadko is a 1 metre telescope located at the Gingin observatory in Western Australia~\cite{Coward:2016jja}. As its FoV is about 0.15 deg$^{2}$, seven tiles are needed to cover the ANTARES point spread function. In 2017, the follow-up was extended to the SVOM/GWAC~\cite{Wei:2016eox} 18 cm telescopes located in China, providing a very large FoV (about 40$^\circ$) but with a not very deep sensitivity (about 15 mag). \\

\begin{figure}[!ht]
\centering
  \includegraphics[width=\linewidth]{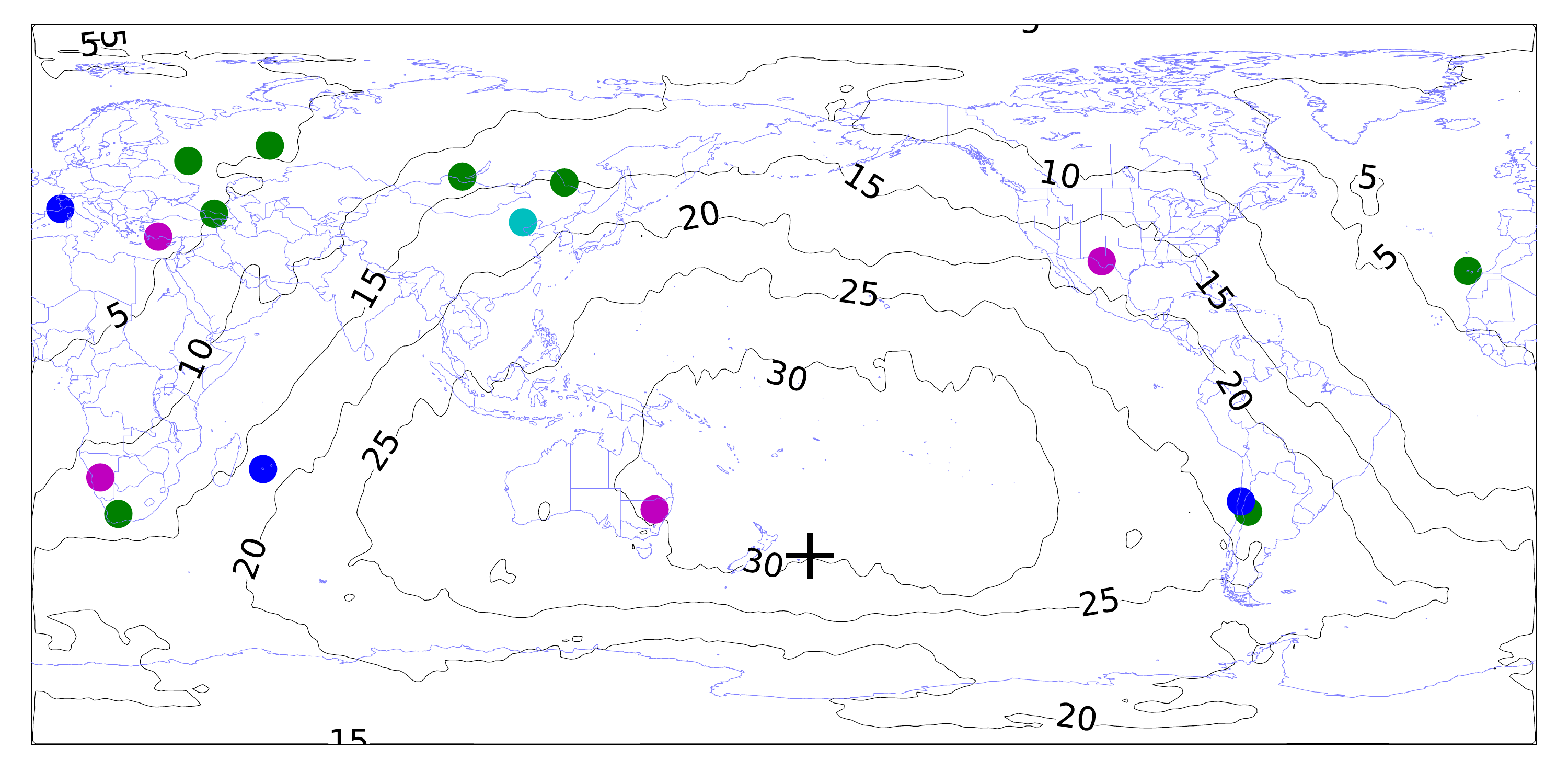} 
  \caption{Follow-up iso-efficiency curves of the ANTARES alerts as a function of the location of a given telescope. The numbers indicate the percentage. The dots represent the positions of MASTER (green), ROTSE (pink), TAROT/Zadko (blue) and SVOM ground telescopes (cyan). The black cross indicates the antipode of the ANTARES location.}
  \label{fig:roboticeff}
\end{figure}

Figure~\ref{fig:roboticeff} shows the probability that an ANTARES alert is followed promptly as a function of the location of a given telescope in the world. This probability map has been computed taking into account the ANTARES alert directions and the observing conditions of the telescopes (night with no bright moon and elevation of the neutrino event greater than 7$^\circ$). As expected, the antipode of the ANTARES location has the best efficiency, but also optical telescopes in the Northern Hemisphere have a significant chance to observe promptly in the direction of the alert. For each alert, the optical observation strategy is composed of an early follow-up (within 24 hours after the neutrino detection), to search for fast transient sources such as GRB afterglows, complemented by several observations during the two following months, to detect for example the rising light curves of CCSN or the flare of an AGN. Each observation is composed of a series of optical images (with clear filter). \\

The Swift satellite~\cite{Gehrels:2004aa} with its X-ray Telescope (XRT~\cite{Burrows:2008ts}) provides a unique opportunity to observe X-ray counterparts to neutrino triggers on account of its large field of view and its very prompt and flexible scheduling processes~\cite{Adrian-Martinez:2015nin}. The detection sensitivity of the XRT is about 5$\times$10$^{-13}$ erg cm$^{-2}$ s$^{-1}$ in 1 ks exposure in an energy band from 0.3 to 10 keV ~\cite{Evans_2013}. Due to the small FoV of the XRT (radius of $\sim$~0.2$^\circ$) and the typical error radius of an ANTARES alert ($\sim$~0.5$^\circ$), each observation of a neutrino trigger is composed of 4 tiles of 2 ks each. This mapping covers about 70\% of the ANTARES point spread function for the very high-energy neutrino trigger. This choice was a compromise between the size of the covered region and the sensitivity of the search. The observation strategy is composed of an automatic response to the neutrino trigger with observations starting as soon as possible. There is an online analysis of the data~\cite{Evans_2013,61ccea70e38a49898265c91a1720cb49} and in case an interesting counterpart candidate is found, further observations are scheduled. \\


The INTEGRAL satellite~\cite{2003A&A...411L...1W} carries a collection of hard X-ray and soft gamma-ray instruments (and a small optical monitor), covering the energy range from 3 keV to 10 MeV. The instruments feature imaging FoV (up to about 1000 deg$^2$) but are also sensitive to gamma-ray emission above $\sim$100 keV from the whole sky. Imaging instruments can be re-pointed to any given target with a delay from a few hours to about 1 day. No reoriented follow-ups of ANTARES events were performed, since none of  them passed at the same time the selection criteria and the observability constraints. However, owing to the wide FoV as well as the exceptionally eccentric orbit of the INTEGRAL spacecraft, the effect of the Earth shadow is negligible, and observations are available at any time the instruments are active (about 85\% of the time). Hence, even if the alert position was not at the centre of the FoV, for every received trigger from ANTARES, INTEGRAL was able to derive a constraining upper limit on impulsive gamma-ray flux, ranging from 2$\times$10$^{-7}$~erg~cm$^{-2}$~s$^{-1}$ to about 6$\times$10$^{-7}$~erg~cm$^{-2}$~s$^{-1}$ for a typical GRB spectrum.\\

Moreover, follow-up observations of a sub-sample of neutrino alerts have also been performed by the Murchinson Wide Field Array (MWA~\cite{2009IEEEP..97.1497L,2013PASA307T}) which is the low frequency ($80-300$ MHz) precursor of the Square Kilometre Array. Its fast re-pointing and its huge field of view (700 deg$^2$ at 150 MHz) is particularly valuable for follow-up of neutrino candidates. The MWA angular resolution is of the order of one arc minute, allowing good localisation of transient radio sources. At the low radio frequencies of the MWA, the dispersion delay implies that the MWA can be re-pointed before the low frequency radio waves arrive at the telescope. This characteristic makes the MWA unique for the follow-up of prompt electromagnetic emission from ANTARES events.\\

A few alerts have also triggered observations by the H.E.S.S. (High Energy Stereoscopic System) imaging atmospheric Cherenkov telescope. H.E.S.S.~\cite{Aharonian:2006pe} has a typical energy threshold of 100 GeV and a field of view of around 5$^\circ$. H.E.S.S. telescopes are located at an elevation of 1800 m above sea level on the Khomas Highland plateau of Namibia. The original array, inaugurated in 2004, is composed of four 12~m diameter telescopes. In 2012, a fifth telescope with a 28~m diameter mirror was commissioned. \\

As the ANTARES data were not public, the collaboration has signed a Memorandum of Understanding (MoU) with each partner to fix the rules of data exchanges as well as the publication and communication of the joint results. \\

\section{Main results}
\begin{figure}[!ht]
\centering
  \includegraphics[width=\linewidth]{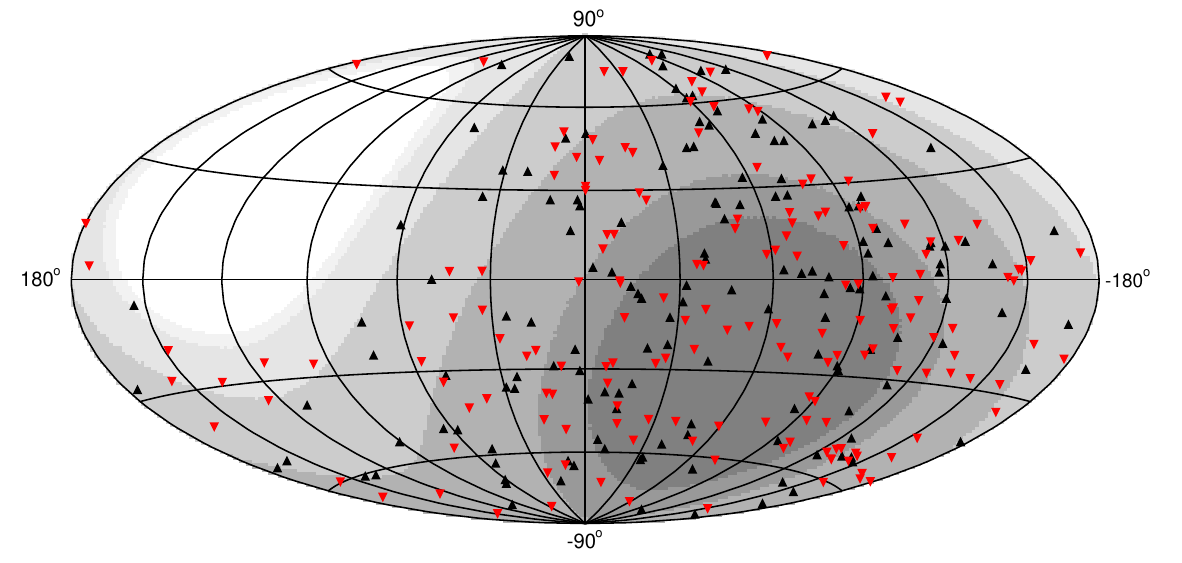} 
  \caption{Sky map in Galactic coordinates showing the directions of all the TAToO alerts: red and black markers correspond to alerts with early follow-up ($<$24 h) and with only late follow-up, respectively. The grey regions indicate the integrated ANTARES visibility, from dark grey, 100\%, to white, 0\%.}
  \label{fig:tatoo_skymap}
\end{figure}

High-energy neutrinos are thought to be produced in several kinds of astrophysical sources, such as GRBs, CCSNs or AGNs. Most of the sources show also transient high-activity phases covering a large range in the time domain, from seconds for the GRB prompt phase to weeks for CCSNs or AGNs.  Between mid 2009 and December 2020, a total of 322 alerts were sent to robotic telescopes. Figure~\ref{fig:tatoo_skymap} shows the directions of the TAToO alerts. 
A total of 26 targets of opportunity were sent to Swift since mid-2013. The typical follow-up efficiency is around 70\%.

\subsection{Results of the prompt follow-ups}
Out of the 322 sent alerts, 218 triggers with an early optical follow-up ($<$24 h after the neutrino time) were analysed (68\% of the sent alerts). Among them, 55 had a delay lower than 1 min (18\%). \\

For example, here are details of the follow-ups performed by MASTER: 187 alerts of ANTARES were followed since 2015. 51 out of 187 alerts were observed starting 1 minute after notice time. 
Longer delays can be explained by several reasons: bad weather on the telescope site, follow-up area under horizon for all telescopes at alert time, coordinates close to the Sun or full Moon, etc. 
\\

\begin{figure}[!ht]
\centering
  \includegraphics[width=0.49\linewidth]{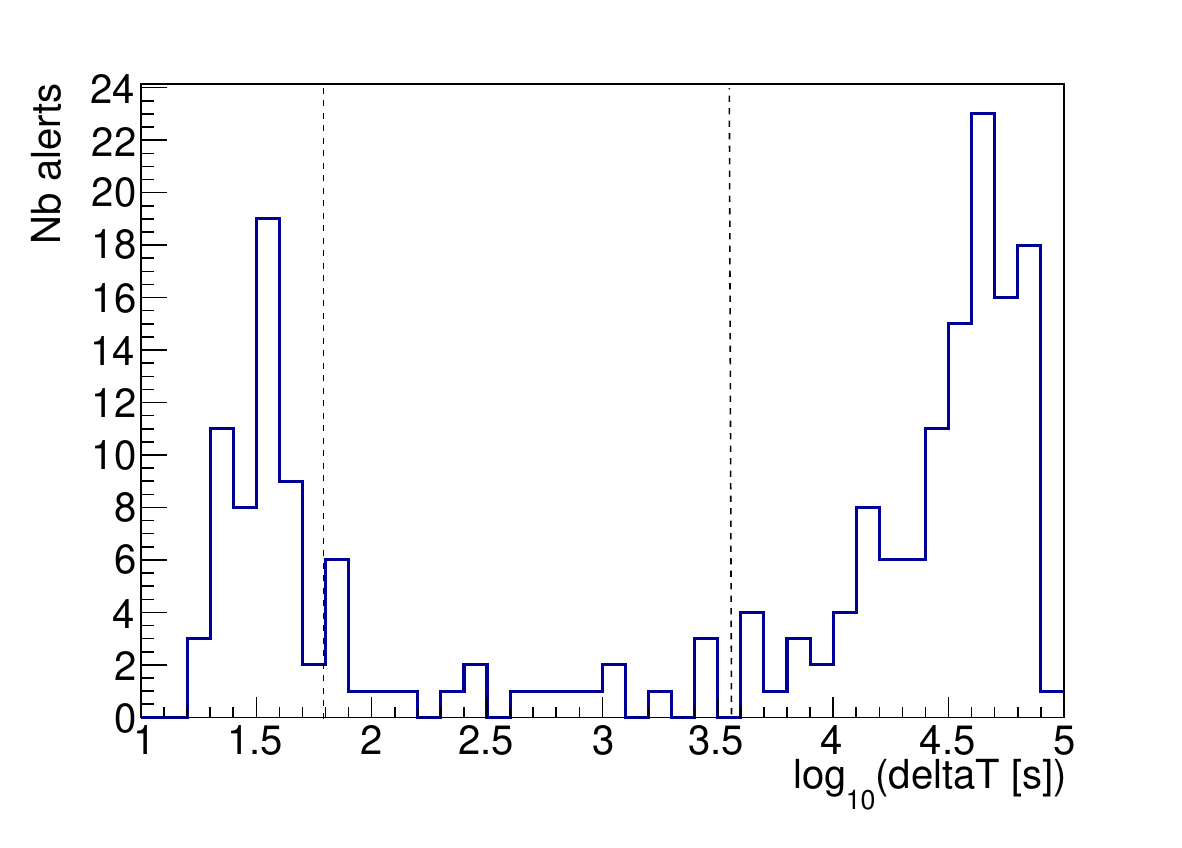}
  \includegraphics[width=0.49\linewidth]{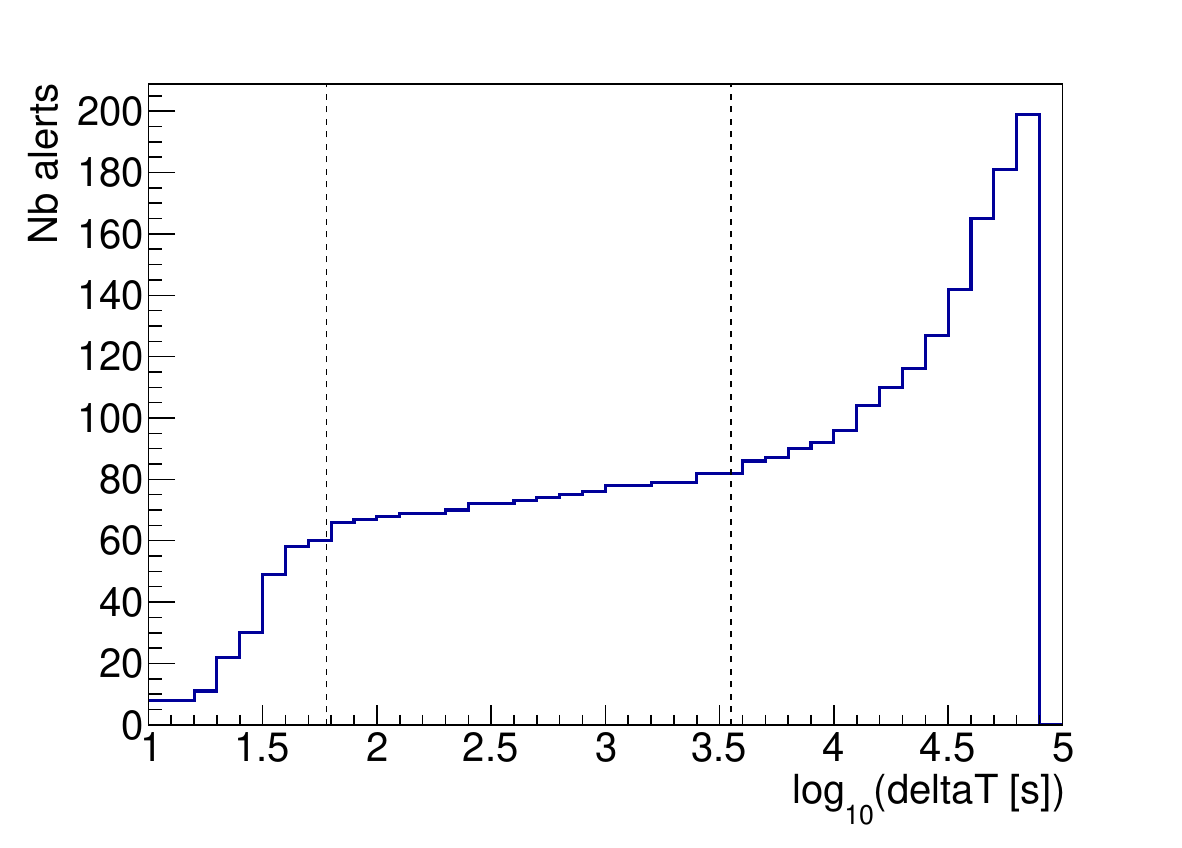}
  \caption{(left) Distribution of the delays between the earliest TAROT or MASTER images of the follow-up and the time of the neutrino (deltaT) for all the ANTARES alerts from 2011 to 2021. (right) Cumulative distribution. The vertical dashed lines indicate a delay of 1 minute and 1 hour.}
  \label{fig:tatoo_delay_image}
\end{figure}

Figure~\ref{fig:tatoo_delay_image} shows the delay between the first image of the follow-up performed by the robotic telescopes of TAROT and MASTER and the time of the neutrino. The minimum delay is 17 s, which includes the time of the alert sending, the transmission of the alert, its reception at the telescope site, the stop of the ongoing acquisition, the pointing of the telescope and the start of a new image. The first bump at a few tens of seconds corresponds to the alerts immediately observable.\\

No clear optical transient counterparts were found and upper limits on the magnitude of a transient astrophysical source (clear filter) were derived~\cite{Adrian-Martinez:2015nin}. Because rapidly-fading sources (in the typical 10--20 min observation duration) were searched for, the signal is supposed to be more important in the first image of the observation, so the upper limits are the limiting magnitude of each first image computed at 5$\sigma$ confidence level (C.L.). Table \ref{tab:OptObs} summarises the characteristics of the optical follow-up with these telescopes for the VHE neutrino alerts. The Galactic extinction of the magnitude, computed as in Ref.~\cite{Schlegel:1997yv}, is also provided.\\

This non-observation could indicate that the follow-ups were made too late, when the sources could have faded below the sensitivity of the optical telescopes, or that the Galactic extinction was too large, or that the sources were located outside the field of view of the telescopes, or that the neutrinos that generated the triggers had another origin. \\

\begin{figure}[!ht]
\centering
  \includegraphics[width=0.7\linewidth]{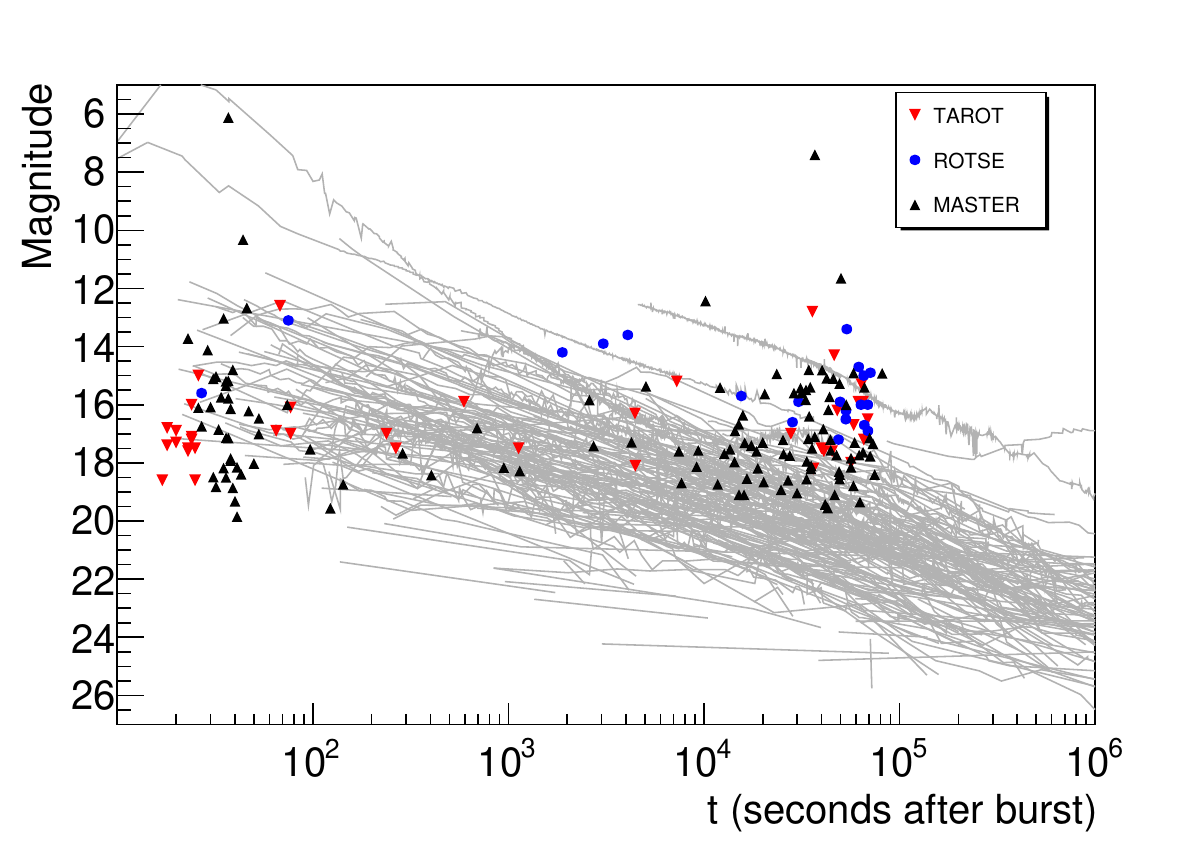} 
  \caption{Comparison between archived optical R band and uncorrected light-curves for 301 GRBs detected during the 1997-2014 period obtained from the GCN circulars (grey lines)  and the magnitude limits inferred for the 208 neutrino alerts during the 2009-2021 period. Red, blue and black markers indicate upper limits on GRB afterglow magnitudes for neutrino alerts set by TAROT, ROTSE and MASTER, respectively.}
  \label{fig:tatoo_grb}
\end{figure}

Assuming the hypothesis that all the neutrinos are issued from GRBs, and by comparing these upper limits with optical afterglow light curves of gamma-ray bursts, it becomes possible to test the GRB association for each neutrino alert. Figure~\ref{fig:tatoo_grb} shows the optical afterglow light curves of GRBs in R band detected from 1997 to 2014 (taken from GCN circulars) and the optical upper limits obtained for each neutrino alert. Note that the comparison between the R and clear observations is not direct, as there is typically less than about 0.5 magnitude difference and depends on the telescope camera and filter. But this small difference does not change the overall conclusion of the studies. \\

The constraints are quite strong ($>$90\%) when the optical follow-up is performed within a few minutes after the neutrino trigger. This probability is computed from the fraction of the detected GRB light curves above the limiting magnitude and the fraction of the neutrino error region covered by the field of view of the telescope. Those constraints on the GRB origin only applied in the case where the EM signal and the neutrino emission happen in the same time window. In this dedicated search, it is not possible to rule out that some neutrino events can be issued from off-axis GRBs, i.e., resulting from GRBs with a jet not pointed towards the Earth (no prompt EM emission). Only the detection of the supernova in time and space coincidence with the neutrino, for the closest events, can constrain this origin (see Section 4.3).\\

\begin{table}[!ht]
\caption{Main results of the optical follow-up by MASTER and TAROT telescopes for the VHE neutrino triggers. Delay refers to the time between the first image of the follow-up and the neutrino detection, Exposure is the exposure of each image, M$_{\rm Lim}$ is the apparent magnitude in R mag and the Galactic extinction is computed according to Ref.~\cite{Schlegel:1997yv}.}
\label{tab:OptObs}
\begin{center}
\begin{tabular}{ccccccc}
\hline
Trigger Id  &   Telescope & Delay  &   Exposure & M$_{\rm Lim}$ & Galactic & Optical \\
            &             & (s)    & (s)       &         &    extinction & transient\\
\hline
ANT130915A & TAROT Chili 	   & 328893   & 180 & 18.1    & 0.09  & 0 \\
ANT130927A  & TAROT Chili     &  227552   & 180 & 17.8    & 0.36 & 0 \\
ANT140123A  & TAROT Calern  & 48025     & 180 & 17.8    & 1.35  & 0 \\
ANT140311A  & TAROT Calern  &  139294   & 180 & 18.8    & 0.07 & 0 \\
ANT141220A  & TAROT Chili      &     131438 & 180 & 18.8   & 0.03 & 0 \\
ANT150109A & No observation & / & / & / & / & / \\
ANT150409A & MASTER SAAO    & 36    &    60 &    18.6    &   0.1	 & 0 \\ 		
ANT150422A  & MASTER Tunka    & 127871    &    60 &     17.3     &    1.3	& 0\\
ANT150809A  & No observation & / & / & / & / & / \\
ANT150901A  & MASTER SAAO    & 35217    &    60 &    20.1     &   1.9		 & 0 \\ 		
ANT151027A  & MASTER Tunka    & 120181    &    60 &    18.2     &    0.2		 & 0 \\ 
ANT151106A  & MASTER SAAO    &  617926    &    60 &    19.4     &   0.2	 	 & 0 \\ 
ANT160227A  & MASTER Amur       & 36    &    60 &    16.0    &    0.7	    	 & 0 \\    
ANT160320A  & MASTER SAAO    & 1840695    &    60 &    18.6    &   0.1		 & 0 \\  
ANT160524A  & MASTER SAAO    &  49446    &    60 &    18.7     &   0.3	 	 & 0 \\ 
ANT170401A  & MASTER SAAO    & 31174    &    60 &    17.5    &   2.1	 		 & 0 \\ 	
ANT170811A  & MASTER OAFA    & 731191    &    60 &    18.4     &   1.5	  	 & 0 \\ 
ANT170902A  & MASTER SAAO    &  34    &    60 &    20.2     &   0.04	 	 & 0 \\
ANT180327A & MASTER SAAO    &  31584    &    60 &    16.6     &   1.0	 	 & 0 \\ 	    ANT180725A & MASTER OAFA    & 10957    &    60 &  16.1    &   12.4     	 & 0 \\ 		  
ANT180917A & MASTER OAFA    & 948    &    60 &  18.2    &   0.05       	 & 0 \\ 	 	
ANT190410A & MASTER SAAO    & 12968    &    60 &  20.5    &   0.03    		 & 0 \\ 		
ANT190428A& MASTER OAFA    & 42447    &    60 &  15.3    &   0.19    	 & 0 \\ 	
ANT191126A & MASTER Tunka    & 43    &    60 &  18.6     &    0.21    		 & 0 \\ 		ANT191231A & MASTER SAAO    &  13598    &    60 &   17.6     &   0.07   	 & 0 \\ 		ANT200108A & MASTER Amur    & 31    &    60 &   15.2     &   0.09     		 & 0 \\ 
ANT201222A & MASTER OAFA    & 143   &   60 &    19.1    &     0.36          & 0 \\
\hline
\end{tabular}
\end{center}
\end{table}

\begin{figure}[!ht]
\centering
  \includegraphics[width=0.7\linewidth]{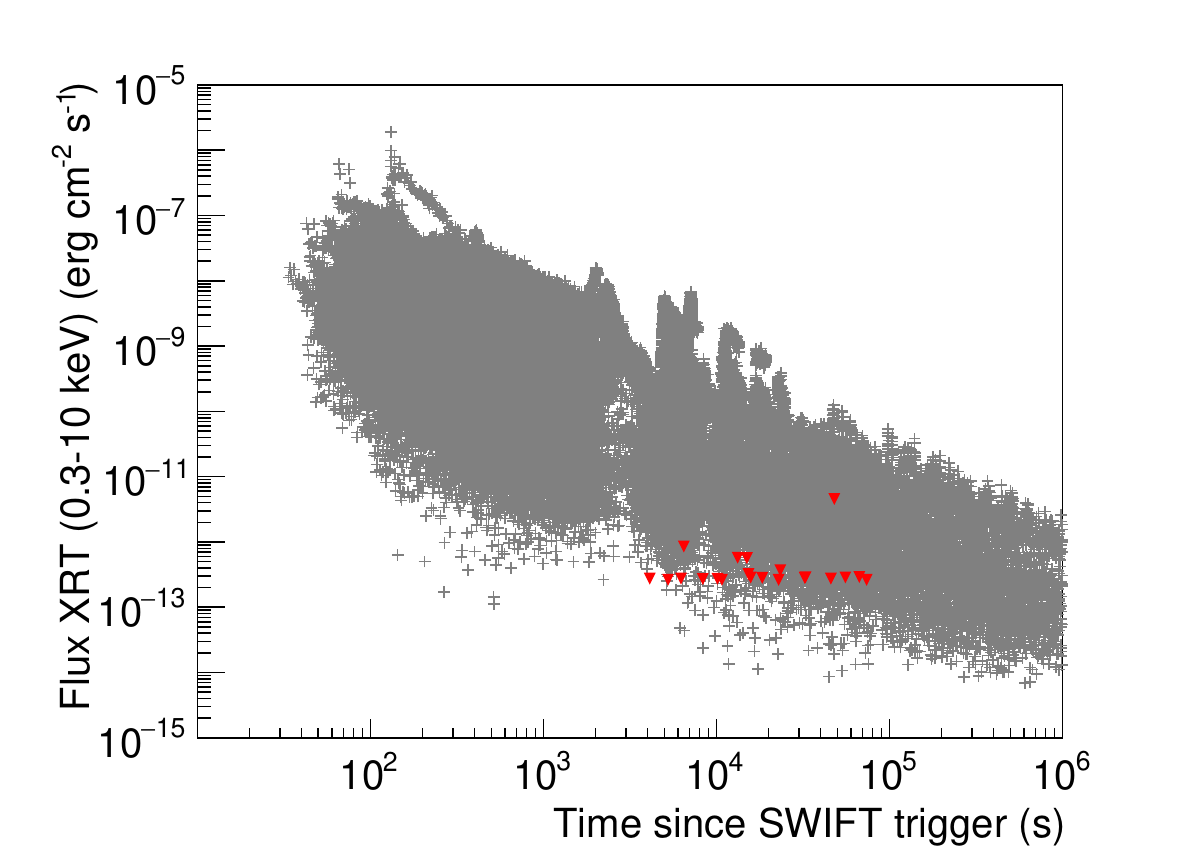} 
  \caption{Comparison between the energy flux of 979 GRB afterglow lightcurves detected in the period 1997-2017 by Swift~\cite{Evans:2007na} (grey crosses) and the flux upper limits obtained for the 18 followed neutrino alerts during the 2013-2020 period (red markers).}
  \label{fig:tatoo_grb_X}
\end{figure}

A similar analysis was carried out with Swift-XRT follow-ups of 19 ANTARES alerts~\cite{Adrian-Martinez:2015nin} (Figure~\ref{fig:tatoo_grb_X}). The average delay of the first Swift observation is around 8 h with a minimum delay of 1.1 h. The probability to reject the presence of an on-axis GRB reaches more than about 70\% for each alert if the X-ray follow-up occurs within a few hours after the trigger. This is mainly dominated by the limited coverage of the four tiles to cover the region of interest defined by the ANTARES angular resolution (see Figure~\ref{fig:psf} left). Table \ref{tab:SwiftObs} summarises the X-ray observations. The particular event ANT150901A with one possible counterpart is described in the following subsection.\\

\begin{table}[ht!]
\caption{
Summary of the result of the Swift-XRT observations of the follow-up of ANTARES alerts (listed in the first column). The following columns contain the delay (in s) between the neutrino event and the first observation; the number of tiles; the total time exposure of the satellite; the number of new sources not catalogued in the ROSAT survey (in parenthesis, the total number of sources identified by the XRT automatic analysis pipeline); the number of counterpart candidates (Count. cand.); and the evaluated 90\% C.L. flux upper limit. For ANT200108A, the observing strategy applied for Swift was not the standard ANTARES one but the IceCube one which covers a larger field of view with 19 tiles but with a smaller exposure (500 s each).}
\label{tab:SwiftObs}
\begin{tabular}{ccccccc}
\hline
Trigger Id  &   Delay  & Nb(tiles) & Exposure  & New sources & Count.  & Flux limit\\
            &    (s)    &     &   (ks)    & (Total sources) & cand. & (10$^{-13}$ erg cm$^{-2}$ s$^{-1}$)\\
\hline
ANT130722A   & 4117     &   4    &      7.3     &	     4 (5) & 0	&   2.7\\
ANT130915A   & 23418  &   4    &      5.5   &     2 (2) & 0	&   3.7 \\
ANT130927A   & 18416   &   4    &      7.1      &     3 (4) & 0	&  2.8 \\
ANT140123A   & 13267   &   4    &      3.5     &     1 (1) & 0	&  5.7 \\
ANT140311A   & 10142    &   3    &      5.5     &    3 (3) & 0	&  2.7 \\
ANT141220A   & 5260     &   4    &      7.6     &    2 (2) & 0	&  2.6 \\
ANT150129A   & 6244     &   4    &      7.5     &    3 (3) & 0	&  2.7 \\
ANT150409A   & 45882  &   4    &      7.3     &    5 (5) & 0	&  2.7 \\
ANT150809A   & 55636  &   4    &      7.0     &    2 (2) & 0	&  2.9 \\
$\mathbf{ANT150901A}$   &  32411	 &   5    &    7.1     &     4 (8) & 1	&  2.8 \\
ANT160227A   & 6483     &   4    &      2.3  &    0 (0) & 0	&  8.5 \\
ANT170401A   & 8344     &   4    &      7.4    &    1 (1) & 0	&  2.7 \\
ANT170811A   & 67025    &   4    &      6.9   &    2 (3) & 0	&  2.9 \\
ANT180327A   & 10765    &   4    &      7.6   &    7 (9) & 0	&  2.6 \\
ANT180725A   & 22901    &   4    &      7.6   &    3 (5) & 0	&  2.6\\
ANT180917A   & 15014     &   4    &      3.5  &    3 (3) & 0	&  5.7 \\
ANT190410A   &  15780	 &   5     &     7.2     &     6 (7) & 0	&  2.9 \\
ANT191126A   & 32786    &   4    &      7.0   &   6 (6) & 0	&  2.9 \\
ANT191231A   & 15337     &   5    &      7.7   &   10 (10) & 0	&  3.2 \\
ANT200108A   & 48025   &   19  &      2.1   &   0 (0) & 0	&  45.5 \\
ANT200127A   & 73856	  &  7     &     13.1   &	        4 (4) & 0	& 2.6 \\ 
ANT201222A   & 43168     & 2     &   2.2   & 0 (0) & 0 & 4.6 \\ 
\hline
\end{tabular}
\end{table}

\subsection{Multi-wavelength follow-up of ANT150901A}
In the X-ray follow-up, a bright and transient counterpart candidate was found for a single neutrino alert: ANT150901A. The associated neutrino had an energy of about 90 TeV with a 1$\sigma$ range between $20 - 300$ TeV. The probability that this neutrino was of cosmic origin is 8\%, using the IceCube definition of the signalness~\cite{2017APh....92...30A} with our measured atmospheric neutrino background and a cosmic diffuse neutrino flux with a spectral index $-2.0$, and a normalisation given by the IceCube diffuse cosmic neutrino flux~\cite{Aartsen:2014gkd}.\\

 Observations with XRT started in September 1$^\mathrm{st}$, 2015 at 16:38:42 UT (namely 9 hours after the neutrino trigger). In the first observations, 8 sources were identified in the field of view. Among them, 5 are catalogued and 3 uncatalogued sources. From this list, one uncatalogued X-ray source has been detected above the Rosat All-Sky Survey (RASS) limit~\cite{Voges:1999ju}, with the flux varying between 5$\times$10$^{-13}$ and 1.4$\times$10$^{-12}$ erg cm$^{-2}$ s$^{-1}$  in the $0.3 - 10.0$~keV band at location: RA = 16h26m2.12s and DEC = $-$27d26m14.8s (J2000) with an uncertainty of 2.4~arcsec (radius, 90\% containment). This source is located at 0.11$^\circ$ from the neutrino direction. As the detected X-ray source seemed to be variable, a GCN circular~\cite{2015GCN.18231....1D} and an Astronomer's Telegram~\cite{2015ATel.7987....1D} were published on September 3$^\mathrm{rd}$, 2015 to encourage further multi-wavelength and multi-messenger observations. \\

\begin{figure}[!ht]
\centering
\includegraphics[width=0.6\textwidth]{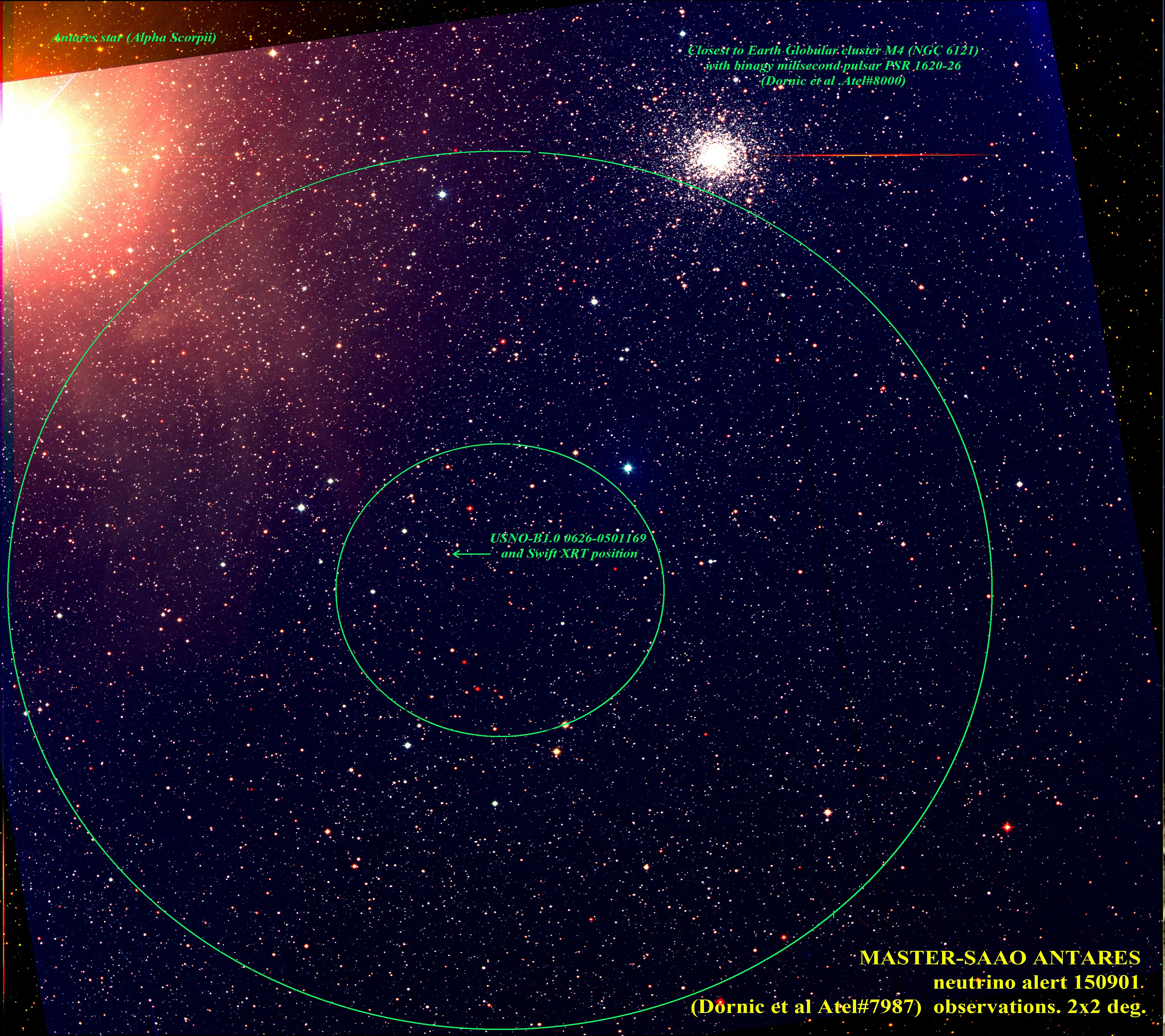}
\caption{Field of view of MASTER corresponding to ANT150901A. The 2 circles have a radius of 0.3 and 0.9 degres. }
\label{fig:fov_master}
\end{figure}

 \begin{figure}[!ht]
\centering
  \includegraphics[width=0.5\linewidth]{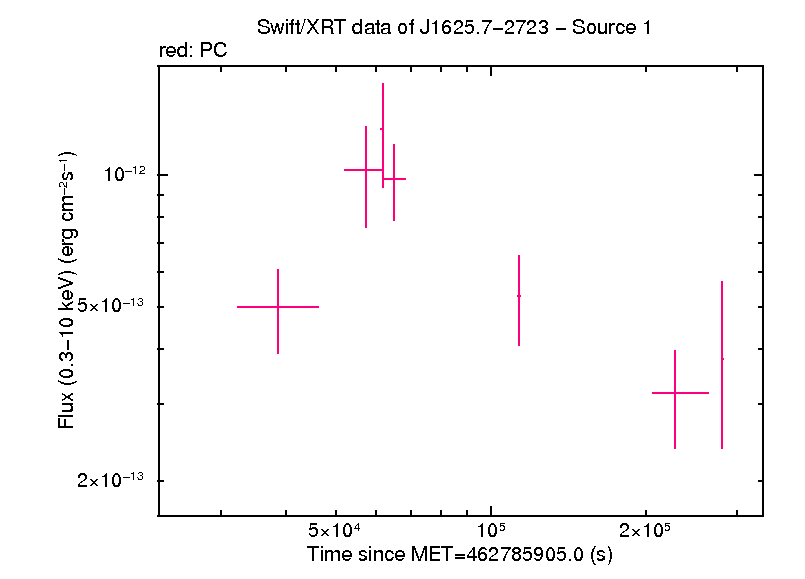} 
  \caption{Light curve measured by the Swift-XRT for the X-ray source identified in the follow-up of ANT150901A. The neutrino detection time is $t = 0$.}
  \label{fig:Xray_LC}
\end{figure}
 
 In parallel, optical follow-ups by the MASTER telescopes~\cite{2015GCN.18240....1D} began 9 hours 45 minutes after the neutrino detection. The location corresponding to the neutrino direction was followed since the first day with two telescopes in South Africa and Canary Island with the R, B, and V filters. This direction was regularly followed by one of the MASTER telescopes during the 8 subsequent days. Table \ref{tab:OptObs} shows the observing conditions and the results of the MASTER follow-up. No optical transient candidate was found in the observations down to a magnitude of 18.67 (60~s exposure). At the position of the X-ray source, MASTER identified a bright star (USNO-B1.0 0626-0501169) of magnitude 12.3 with a light curve showing no flux nor color variations just after the time of the alert~\cite{2015ATel.8000....1D}. Figure~\ref{fig:fov_master} illustrates the field of view of MASTER. The globular cluster M4 is 0.97$^{\circ}$ away and the Antares star (Alpha Scorpii) is at a distance of 1.2$^{\circ}$ away from the ANT150901A neutrino direction.\\
 
 To characterise the bright star USNO-B1.0 0626-0501169 and to test the association between the X-ray flare and this bright star, further observations were requested via a GCN circular. Additional observations with XRT showed a flare with a characteristic length of around two days (Figure~\ref{fig:Xray_LC}) just after the neutrino detection. \\

Moreover, there are no detected galaxies within about 5~arc sec from the Swift X-ray source in the Pan-STARRS PS1 catalogue~\cite{2015ATel.8027....1S}. The bright star was saturated and there were several faint sources in the vicinity of the point spread function but none of them looked like extended galaxies. Follow-up data which allowed the nature of Swift J1625.7-2723 to be constrained are described in the next paragraphs. A total of 19 multi-wavelength observatories answered to this trigger covering the full EM spectrum: one radio telescope, 11 optical/IR telescopes, four X-ray/gamma-ray satellites and four very high-energy gamma-ray observatories. \\

The IceCube Collaboration performed a follow-up analysis and did not find any neutrino candidate with reconstructed energy above 10 TeV during a $\pm$ 1 day time-window centred on the ANTARES event time~\cite{2015ATel.8097....1K}. Fermi/GBM~\cite{2015GCN.18352....1B}, INTEGRAL~\cite{2015ATel.7995....1F} and MAXI/GSC~\cite{2015ATel.8003....1N} performed a high-energy follow-up quickly after the alert message that yielded no high-energy counterpart. A search for a gamma-ray counterpart was carried out using the archive Fermi/LAT Pass8 data, applying a standard unbinned likelihood analysis using the LAT analysis tools (v10r0p5). All photons (event class \texttt{P8R2\_SOURCE\_V6}) within a region of interest of 15$^\circ$ radius were selected within the energy range [100 MeV; 500 GeV]. For all the sources present in catalogues, their fluxes were fixed to the values in the catalogue while the normalisation of the Galactic and diffuse isotropic background were left free in the likelihood maximisation. Four different time windows were considered in the analysis from $\pm$ 6 hours to $\pm$ 6 days. No significant counterpart was detected and 95\% C.L. upper limits on the energy flux have been computed assuming a point-like source with a fixed spectral index of -2.0. Results are given in Table \ref{tab:fermi_UL}.\\

\begin{table}[!ht]
\caption{Fermi-LAT 95\% C.L. energy flux upper limit in the 100 MeV to 500 GeV energy range from the sky position of neutrino event ANT150901A.}
\label{tab:fermi_UL}
\centering
\begin{tabular}{lcc}
\hline
Time window & 95\% C.L. upper limit\\
 & MeV~cm$^{-2}$~s$^{-1}$\\
\hline
$\pm$~6~h  & 2.2$\times$10$^{-4}$\\
 $\pm$~12~h & 1.5$\times$10$^{-4}$\\
 $\pm$~24~h  & 5.0$\times$10$^{-5}$ \\
 $\pm$~6~d  & 7.0$\times$10$^{-6}$ \\
\hline
\end{tabular}
\end{table}

At very high energies, the H.E.S.S. telescope was triggered directly by the neutrino alert and the follow-up started on September 1st, at 18:58 UT as soon as good observation conditions were reached~\cite{2017ICRC...35..653S}. No source was detected in the H.E.S.S. FoV. Consequently an upper limit on the gamma-ray flux was derived as $\Phi$(E$>$320$\mathrm{~GeV}$)$<$ 2.4$\times$10$^{-7}$~m$^{-2}$~s$^{-1}$ (99\% C.L.). The MAGIC telescopes performed follow-up observations of a $\sim$~3$^\circ$ diameter sky region centred at the proposed X-ray counterpart of the ANTARES-detected neutrino event~\cite{2015ATel.8203....1M}. MAGIC observed the source under non-optimal conditions due to the large zenith angle ($>$~60$^\circ$), starting from September 3$^{\mathrm{rd}}$ at 20:54 UT, 2.5 days after the ANTARES detection. Observations were carried out for 6 nights, ending on September 8$^{\mathrm{th}}$, and collected a total on-source exposure of 4 hours. No significant emission from the location of the X-ray source was detected.  \\

\begin{figure}[!ht]
\centering
\includegraphics[width=0.6\textwidth]{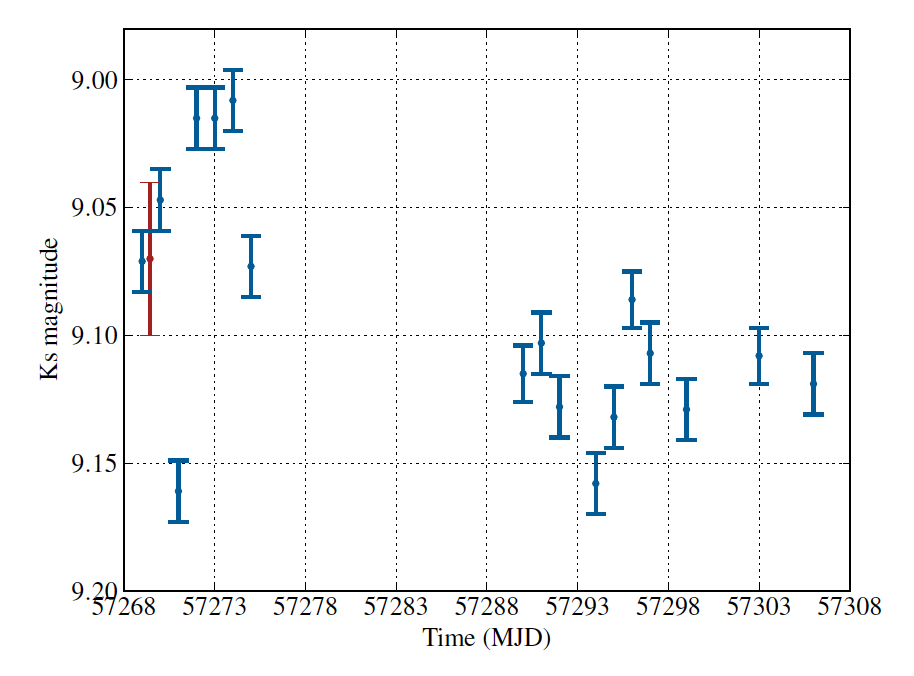}
\caption{Near infrared light curve of USNO-B1.0~0626-0501169 (Ks-band) starting $\sim$2 days after the ANTARES trigger (red point: ATel \#8006).}
\label{fig:nir_LC}
\end{figure}

Long-term near-infrared (NIR) Ks-band observations of USNO-B1.0~0626-0501169 were performed using the Infra-Red Imaging System (IRIS~\cite{Hodapp2010}) located at the observatory Rolf Chini Cerro Murphy (Chile). The resulting light curve is shown in Figure~\ref{fig:nir_LC} together with the Ks-band magnitude (red point) provided in Ref.~\cite{2015ATel.8006....1T}. A NIR flux enhancement is observed at least until September 10$^{\mathrm{th}}$, while a significant decrease of the flux is seen on September 16$^{\mathrm{th}}$. This significant  variation of $\sim$~0.11 mag, followed by a decrease of  $\sim$~0.15 mag, could be interpreted as a succession of two NIR flares. Unfortunately, no later data were available, which prevented a precise constraint on the NIR flux evolution. 

\begin{figure}[!ht]
\centering
\includegraphics[width=0.6\textwidth]{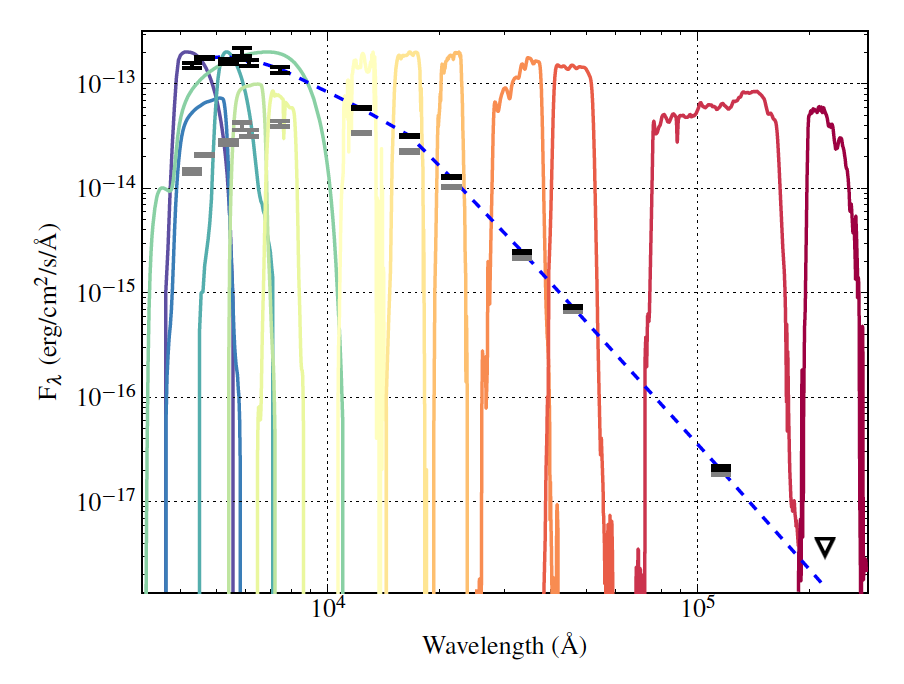}
\caption{Spectral energy distribution (black dots) and Kurucz stellar model fit (dash blue). Grey dots represent the photometric data before correction of the line-of-sight extinction. The different wavelength bands are represented in colours. Downward triangle stands for an upper limit.}
\label{fig:SED}
\end{figure}

Optical and infrared archival data were used to build the spectral energy distribution (SED) of USNO-B1.0 0626-0501169. Although not contemporaneous, those multi-wavelength data are considered as acquired during a similar persistent flux period and any flux variation due to flaring behaviour is considered as small as the typical flux uncertainties. The SED was then fitted by the Kurucz stellar model (ATLAS9~\cite{1992IAUS.149.225K}). The result is shown in Figure~\ref{fig:SED}. The best fit gives A$_V$ = 1.405~$\pm$~0.038, T$_\mathrm{eff}$ = 4750~$\pm$~125~K, log(g) = 4.50~$\pm$~0.59, an age of 19~Myrs, a mass of 1.5~M$_{\odot}$ and a radius of 1.1~R$_{\odot}$. Those parameters are consistent with both a G$-$K star and a RS~CVn primary component. Other multi-wavelength observations by NOT~\cite{2015ATel.7994....1D}, SALT~\cite{2015ATel.7993....1D}, CAHA~\cite{2015ATel.7998....1C}, Wifes on the ANU telescope~\cite{2015ATel.7996....1T} and in radio  (VLA Jansky ATel ~\cite{2015ATel.7999....1H, 2015ATel.8034....1T}) confirm this source classification.

These observations point to USNO-B1.0 0626-0501169 being a young accreting G$-$K star or a binary system of chromospheric active stars (RS CVn), undergoing a flaring episode that produced the X-ray emission. This classification is confirmed by 300 -- 2480\,nm medium-resolution spectra acquired with the ESO VLT/X-Shooter spectrograph which shows the presence of H$\alpha$, H$\beta$, CaII, NaID, Fe+Li and CaI lines, typical of G--K stars. Therefore, this source seems not to be the origin of the bright ANT150901A neutrino with a probability of 3\% of chance association. This probability has been computed taking into account the large-scale spatial distribution of X-ray active stars to estimate the average number of active stars in a given direction using ROSAT and Swift catalogues, the flaring rate (proportion of time spent in an active state) and the position and the error region of the neutrino ANT150901A.

\subsection{Results of the long term follow-ups}
Core-collapse supernova, with rising light-curve in the weeks following the neutrino observation, and flares of AGNs, were the two main sources for the study of the long-term follow-up. Among the 322 sent alerts, 224 had sufficiently good optical long-term follow-ups, i.e., at least 3 (2) nights of observations from the TAROT (MASTER) network. Among these 224 alerts, 77 were triggered by the directional trigger and 153 by the single HE/VHE triggers. No significant slowly varying transient optical counterparts were found in association with a neutrino trigger. The expected number of accidental SN detections, i.e., a CCSN detection in coincidence with a background neutrino event, was estimated to be 0.4 for the 224 alerts assuming a rate of 1 CCSN per year within a sphere of 10 Mpc (i.e., 2.4$\times$10$^{-4}$ yr$^{-1}$ Mpc$^{-3}$~\cite{2005PhRvL..95q1101A}). Our result of no optical long-term follow-up due to a CCSN in correspondence with 224 neutrino alerts is consistent with the small expectated CCSN number with a probability of $\sim$0.7. Note that there are some CCSN detected in the fields of view of the alerts but the neutrino times are not compatible with an estimated explosion time.

\subsection{Search for the origins of the neutrino alerts}
As no GRB and no CCSN have been identified as a counterpart of any of the neutrino alerts, the search was extended to other types of high-energy transients. A number of IceCube neutrino alerts have been potentially associated to blazars. The most famous example is the association between the high-energy neutrino IC170922 and the flaring blazar TXS 0506+056~\cite{IceCube:2018dnn}. The discovery of an optical changing state in TXS 0506+056 by MASTER related to the neutrino ~\cite{2020ApJ...896L..19L} is particularly interesting, at a 50$\sigma$ significance level. This blazar was detected by MASTER to be in the off-state after one minute and then switched to the on-state no later than two hours after the event. \\

\subsubsection{Correlation between the ANTARES neutrino alerts and the BZCAT catalogue}
For each neutrino alert, correlations between the neutrino directions and the blazars listed in the Roma-BZCAT catalogue~\cite{Massaro:2008ye, Massaro:2015nia} were investigated. This blazar catalogue contains 3561 sources, compiling the results of multi-frequence observations. Correlations between sources in the catalogue and neutrino candidates were searched for by applying a cut on the angular distance between the neutrino direction and the blazar coordinates (to be lower than 0.7$^\circ$) and requiring that the MASTER optical telescope observed the field not later than 2 hours after the neutrino detection time.
After the filtering, four blazars were selected. The ANTARES events and the blazar name in the catalogue are listed in Table~\ref{tab:blazars_bzcat_neutrinos}.\\

\begin{table}[!ht]
\caption{List of the four blazars of the Roma-BZCAT correlated with an ANTARES alert. See text for the definition of correlation.\\}
\label{tab:blazars_bzcat_neutrinos}
\centering
\begin{tabular}{lcc}
\hline
Alert name & Blazar name\\
\hline
ANT160815A	& PKS 1806-458 \\
ANT181108A	& RXS J22562-3303 \\
ANT190225A	& PKS 0341-256 \\
ANT190315A	& Centaurus A \\
\hline
\end{tabular}
\end{table}

\begin{figure}[!ht]
\centering
\includegraphics[width=\textwidth]{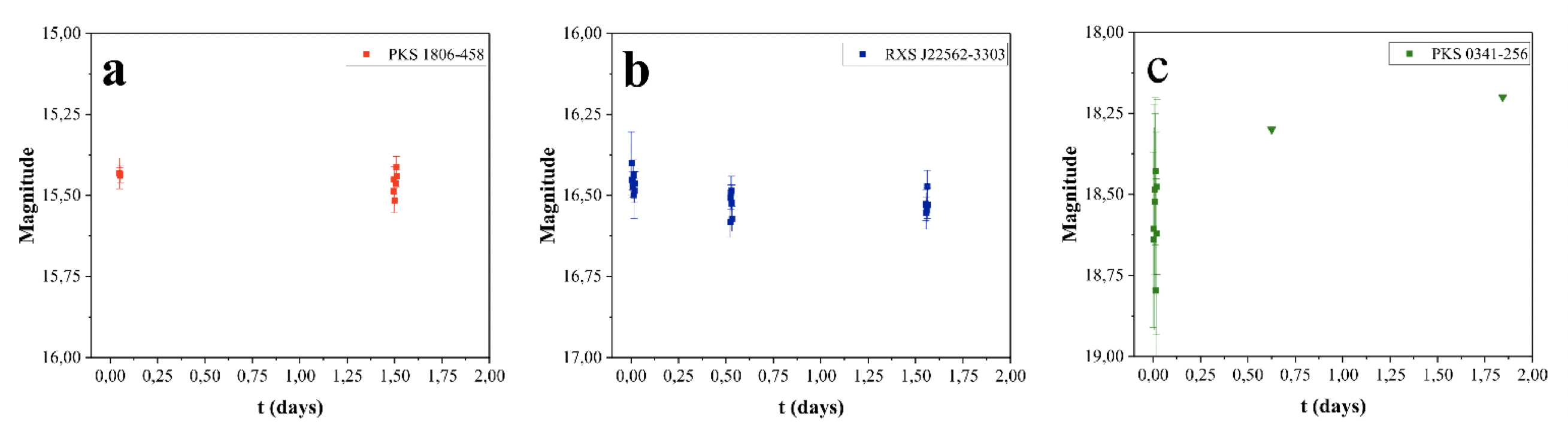}
\caption{Light curves of blazars PKS 1806-458 (a), RXS J22562-3303 (b) and PKS 0341-256 (c) located in the vicinity of ANT160815A, ANT181108A and ANT190225A, respectively. Downward triangles stand for upper limits. The time $t~=~0$ corresponds to the arrival time of the neutrino.}
\label{fig:LC_bzcatbl}
\end{figure}

For each blazar, the MASTER Observatory built a light curve up to 2 days after the neutrino time. The photometry has been computed using a 8-sec aperture and the Gaia EDR3 catalog~\cite{2021A&A...649A...1G} for reference. No significant (more than 3$\sigma$) flux variation in the time interval was found. Figure~\ref{fig:LC_bzcatbl} shows the corresponding light curves of the blazar PKS 1806-458, RXS J22562-3303 and PKS 0341-256. In the case of Centaurus A (a close radiogalaxy), due to the extension of the source, a search for new transients has been performed rather than performing the photometry. No new transients have been detected down to 18 mag. Therefore, these AGNs are not considered as being likely sources of the neutrino candidates. \\

\subsubsection{Search for optical flux variation of blazars observed with MASTER and Gaia data}
In order to study the optical variability of blazars that coincides with the uncertainty area of the incident HE and VHE ANTARES neutrino alerts, their light curves were studied in the database of MASTER telescopes for the period from 2004 to 2020. Also, in the search for variability, information from the database of the Gaia satellite was used\footnote{\url{http://gsaweb.ast.cam.ac.uk/alerts/home}}.

\begin{figure}[!ht]
\centering
\includegraphics[width=0.45\textwidth]{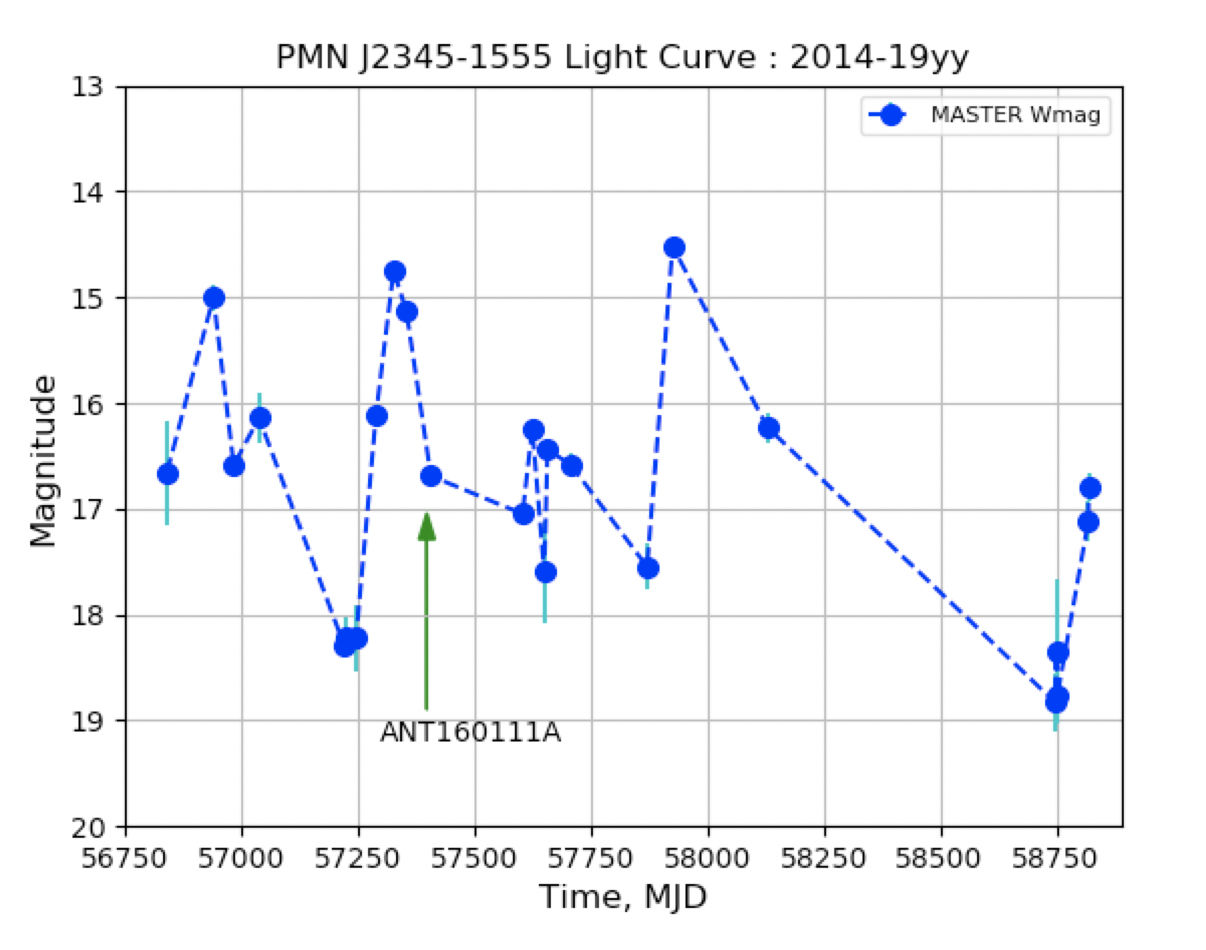}
\includegraphics[width=0.45\textwidth]{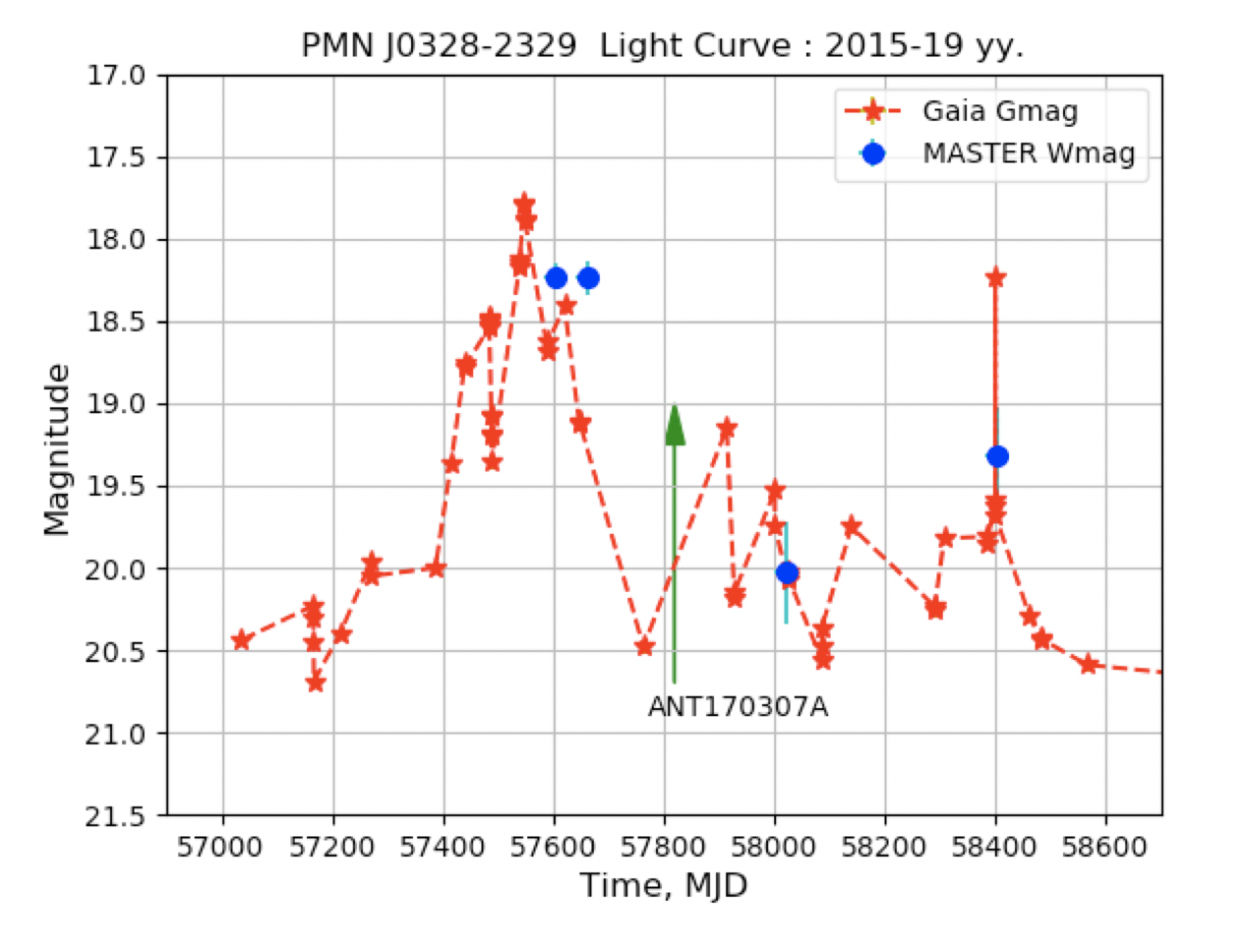}
\includegraphics[width=0.45\textwidth]{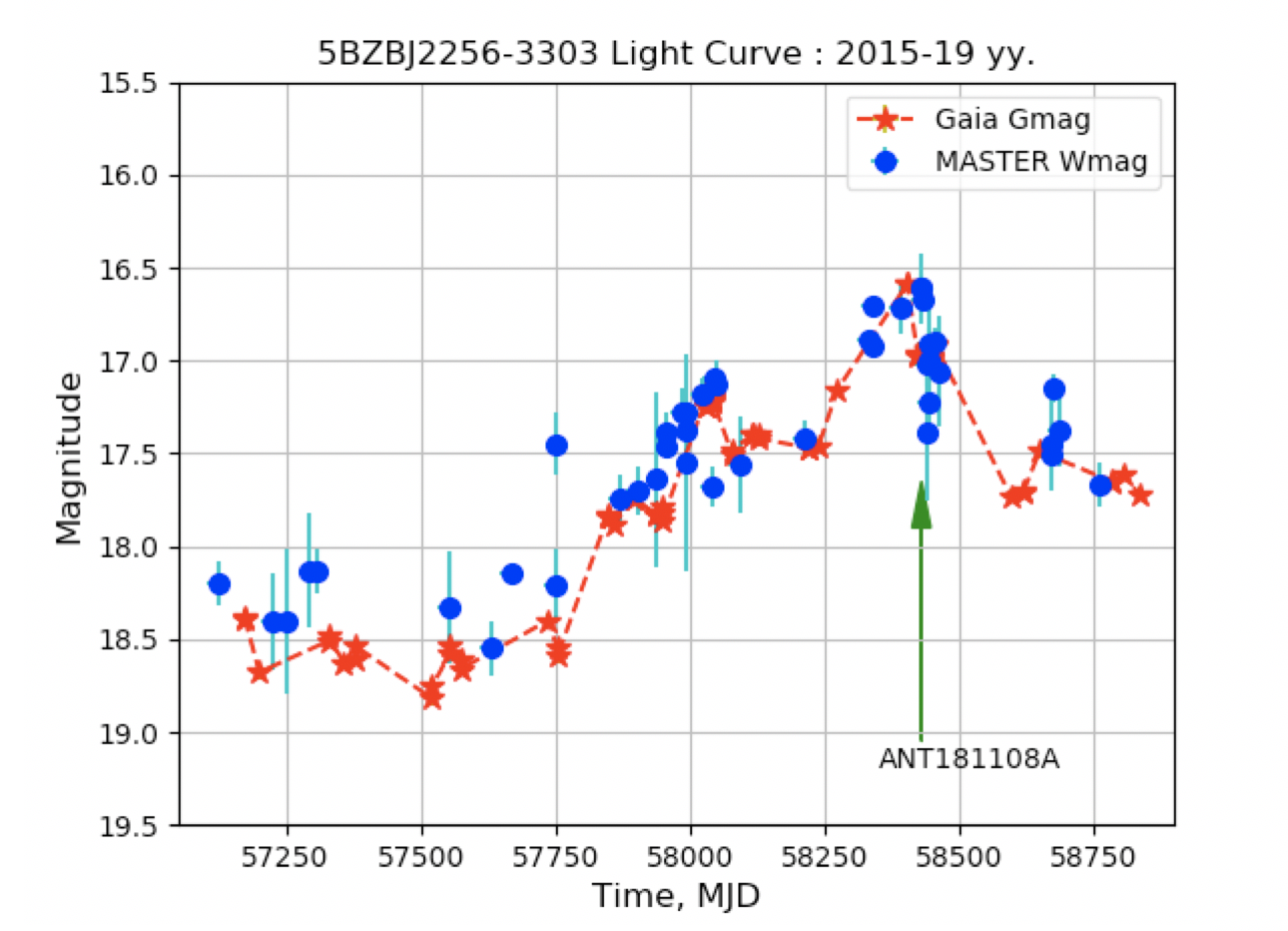}
\caption{Detailed light curve of the blazars PMN J2345-1555 (top left), PMN J0328-2329 (top right) and 5BZB J2256-3303 (bottom) observed by MASTER (W mag) and Gaia (G mag). The arrows indicate the time of the HE neutrino events. }
\label{fig:LC_blazar_master_g}
\end{figure}

Of the 179 neutrino alerts received by the MASTER observatory and whose arrival areas were surveyed, only 20 high energy neutrino events contained blazars in the error-box arrival with a radius smaller than 0.7$^\circ$ (see Table~\ref{tab:blazars_neutrinos}). Significant optical variability of three blazars out of 20 possible HE neutrino events was found. The light curves of these three blazars are shown in Figure~\ref{fig:LC_blazar_master_g}. However, confirming the potential association between HE neutrinos and optical flaring blazars will necessitate the observation of additional neutrino events coincident with blazar optical flares and a thorough theoretical assessment of the underlining physical mechanisms driving such an association.\\



\begin{table}[!ht]
\caption{Results of the study of the optical variability of blazars for which their positions are compatible with the neutrino directions. Dist refers to the distance in degrees between the blazar and the neutrino coordinates.  z is for the redshift. Mag is the magnitude, the letter at the beginning indicates the filter. Var stands for variability.\\}
\label{tab:blazars_neutrinos}
\centering
\begin{tabular}{cccccc}
\hline
Alert & Dist (deg) & 	 Blazar  &	 z & 	Mag &	 Var\\
 \hline
\multirow{3}{*}{ANT150517A} & 0.356 & 2E 0336-2453 & 0.251 & R18.6 & / \\
                            & 0.276 & ESO 482-14 & 0.044 & R14.2 & / \\
                            & 0.342 & QSO B0336-2454 & 0.043 & V17.4 & / \\
\hline
ANT150518A	& 0.649 &	5BZQ J2003-3251 &	 3.773 &	 R17.1 &	 / \\
\hline
ANT160111A	& 0.610	& PMN J2345-1555 &	 0.621	 & R18.5	& Yes\\
\hline
ANT160210A	& 0.463	& PKS 0325-222	& 2.220	& G18.9	 & / \\
\hline
ANT160731A	& 0.426	& QSO J0724-0715 &	 0.271	& V18.0	 & / \\
\hline
ANT160815A	& 0.327	 &PKS J1809-4552 &	 0.070	&  G17.6	 & / \\
\hline
\multirow{3}{*}{ANT160902A} &  0.481 & NVSS J102827+055515 & 0.234 & V19.6 & / \\
                            & 0.683 & SDSS J10279+0631 &  0.158 & V18.7 & / \\
                            & 0.434	& SDSS J10285+0600 & 0.313	& G18.3 & / \\ 
\hline
ANT161031A	& 0.280	& PKS 0524+034 & 	 0.509	 & G19.4	 & / \\
\hline
ANT161209A	&  0.614	& PMN J2256-6533	 & 0.247	 & G16.8 & / \\
\hline
ANT161214A	 & 0.381	& 5BZU J1928-0456	 & 0.587	 & V18.4	 & / \\
\hline
\multirow{3}{*}{ANT170307A} &  0.358 & QSO B0327-241 & 0.895 & V18.1 & / \\
                            & 0.126 & QSO HE0327-2348 & 1.550 & V17.6 & / \\
                            & 0.491	& PMN J0328-2329 & / & V18.6 & Yes \\
\hline
ANT170406A	&  0.395 & 	PKS 0420-484	&  0.527	 & V17.5	 & / \\
\hline
ANT170907A	 & 0.575	& PKS 2053-323	 & /	 & G18.8	 & / \\
\hline
ANT180526A	 & 0.310	& QSO B1925-610	 & 3.254	 & R19.9	 & / \\
\hline
\multirow{3}{*}{ANT180608A} &  0.597 & GB6 J1231+1421 & 0.256 & G17.7 & / \\
                            & 0.441 & NGC 4501 & 0.007 & V13.2  & / \\
                            & 0.561 & NGC 4548 & 0.002 & V13.6 & / \\
                            & 0.308 & QSO B1230+1440 & 0.313 & G14.4 & / \\
                            & 0.482 & QSO B1230+1430 &  0.332 & G17.2  & / \\	
\hline
ANT181108A	& 0.686	& 5BZB J2256-3303 &	 0.243	& G18.5 	 & Yes \\
\hline
ANT190225A	& 0.202	& PKS 0341-256	&  1.419 &	 G18.8	& / \\
\hline
ANT190315A	& 0.402	& 5BZU J1325-4301  &	 0.002	& V6.8	& / \\
\hline
\multirow{2}{*}{ANT191011A} &  0.048 & QSO B0317+185 &  0.190 & G17.9  & / \\
                            & 0.217	& PKS J0319+1901 & 0.296 & G20.3  &	 / \\
\hline
ANT201219A	& 0.309	& 5BZG J1154+1225 &	 0.081 &	 G18.8  &	  / \\
\hline
\end{tabular}
\end{table}

\subsection{Results of the radio follow-up}
A search for radio counterparts of two candidate high-energy neutrino events detected in November 2013 (ANT131121A) and March 2014 (ANT140323A) was performed using the Murchison Widefield Array~\cite{Croft:2016lhf}. Each neutrino candidate had arrival direction consistent ($\leq$ 0.5$^\circ$) with the positions of two galaxies within 20~Mpc from Earth: NGC1374 and ESO358-015 for ANT131121A, and ESO499-037 and PGC29194 for ANT140323A. PGC29194 (the Antlia Dwarf Galaxy), at a distance of 1.3 Mpc, is located just 0.1$^\circ$ from the neutrino direction. 
An optical follow-up was also performed for both neutrino events. For ANT131121A, 12 series of observations containing 6 images each were performed with the TAROT telescope in Chile from 2 to 61 days after the trigger. No optical transient was identified to a limiting magnitude of about 19 (clear filter). For ANT140323A, 8 images were taken with ROTSE 3b in Texas (starting $\sim$ 15 h after the trigger) and 10 images with TAROT Chile up to 45 days after the trigger, according to the long-term strategy. No transient counterpart was found to limiting magnitudes of 16.4 (prompt) and 18.7 (long-term).\\

No strongly varying radio counterpart was identified for the two triggers. Using 5$\sigma$ upper limits of $90 - 340$~mJy, the upper limit for low-frequency radio emission was set to 10$^{37}$ erg s$^{-1}$ (at 150~MHz) for progenitors located at 20~Mpc. These limits do not strongly constrain the late-time emission from even the most luminous radio supernovae or GRBs at these distances. Optical limits are more stringent for these distances. Neither trigger was optimally placed within the MWA field of view: ANT131121A was $\sim$~8$^\circ$ from the pointing centre of the observation, and ANT140323A was $\sim$~17$^\circ$ away. Particularly in the latter case, the fall-off in primary beam response means that noise in the region of the image near the trigger position is higher than ideal. If the neutrino signal is produced not in nearby galaxies but originates in binary neutron star mergers, the limits constrain the progenitors to be at a redshift z $\geq$ 0.2 ($\sim$1~Gpc)~\cite{Croft:2016lhf}. \\

ANTARES also sent a sub-sample of the neutrino alerts in real time to MWA. MWA is electronically triggered, allowing rapid follow-up observations within 8\,s of any event occurring above its elevation limit of $30^{\circ}$~\cite{2019PASA...36...46H} to be performed. Given its location in the Southern Hemisphere, this corresponds to approximately 30\% of upgoing events detected by ANTARES. A fast imaging pipeline was developed to search for transient emissions~\cite{2019A&C....27...23H}, including fast radio bursts~\cite{2018ApJ...867L..12S}, and any other rapid transients on timescales down to 5\,s~\cite{2017MNRAS.466.1944M}. The MWA's all-sky survey GLEAM~\cite{2017MNRAS.464.1146H} also provides a catalogue to search for radio sources with enhanced emission on longer timescales. Using this system, 5 ANTARES events were triggered, with 30 minutes of data captured for each of them. \\

\subsection{Results of the H.E.S.S. follow-up}
In 2016, ANTARES and H.E.S.S. Collaborations signed an MoU to exchange information and alerts in order to exploit the intimate connection between high-energy neutrinos and very high-energy gamma rays. The H.E.S.S. telescopes have followed two ANTARES alerts shortly after the neutrino detection: ANT150901A and ANT170130A. For ANT150901A, the observations started on September 3$^{\mathrm{rd}}$, 2015, at 18:58 UT (11 hours and 20 minutes after the neutrino time) as soon as the necessary observation conditions were reached. No very high-energy gamma-ray source was identified over the 1.5 h observations. This translates into an upper limit on the gamma-ray flux of $\Phi$($E~>~320$~GeV) $<$ 2.7 $\times$ 10$^{-8}$ m$^{-2}$ s$^{-1}$ at 99\% C.L.~\cite{2017AIPC.1792f0006S}. The neutrino ANT170130A direction was also followed by H.E.S.S. with a very short delay of 32 s during 1 hour and again over 45 min the night after. The preliminary analysis revealed no source detection in the neutrino field of view~\cite{2017ICRC...35..653S}. 


\section{Conclusions}
Real-time multi-messenger campaigns are crucial in unveiling
the sources of the most energetic particles and the acceleration mechanisms at work. Neutrinos would provide insights into the physics of stellar explosions, compact object mergers, and relativistic jets, as well as particle acceleration processes. Multi-messenger campaigns rely on the quasi-online communication of potentially interesting observations to partner instruments, with latency of a few minutes, at most. Such alerts are the only way to achieve simultaneous observations of transient phenomena by pointing instruments.\\

This legacy paper summarises more than ten years of follow-ups of the ANTARES high-energy neutrino alerts. This is a very promising method to identify transient phenomena as neutrino production sources, since it is unlikely that the detection of one counterpart associated in time and direction with one neutrino will happen fortuitously. The ANTARES dedicated alert system, TAToO, was able to send alerts to the external community within $\sim$6 seconds after the time of the neutrino detection and reconstructed ANTARES events with an angular accuracy better than 0.5$^\circ$. The triggers were followed by several multi-wavelength facilities such as robotic optical telescopes located all around the world, a radio telescope, two X-ray/gamma-ray satellites and one very high-energy gamma-ray telescope. More than 300 alerts were sent and no counterparts associated with neutrino candidates were significantly detected in the different searches. \\


A decade ago, GRBs were very popular candidates for neutrino production. Nowadays, offline studies by IceCube~\cite{IceCube:2016ipa} and ANTARES~\cite{ANTARES:2020dpd} have set stringent constraints on the GRB/neutrino association. Detected GRBs can only contribute to a few percents of the neutrino cosmic diffuse flux. Note that this does not rule out low-luminosity GRBs (LLGRBs) to have a larger contribution. Due to the limited sensitivities of the current gamma-ray satellites, this LLGRB population is essentially unconstrained. The future GRB satellite SVOM will provide greater sensitivity to LLGRBs~\cite{Wei:2016eox}. For the CCSNe, the main science case relies on hidden jets inside the supernovae~\cite{Senno:2015tsn, Murase:2013ffa}. This model is quite popular since it can establish the link between the long GRB and the CCSNe populations. The non-observation of supernovae correlated in time with neutrinos rules out most of the parameter space of this model. \\

Precision is key to obtain good follow-ups. Therefore, it is important to achieve the best angular resolution in real time and a good control of the absolute pointing accuracy. Moreover, it is highly desirable to be able to properly reconstruct in real time all event topologies spread over a very large energy range, not only restricting to the high-energy muon neutrino channel. Even if the angular resolution for the other event topologies (electron and tau neutrinos produce mostly shower-like events) is limited, these channels are particularly interesting since the atmospheric background contamination is very low. In this respect, KM3NeT~\cite{KM3Net:2016zxf} is starting to enter into the multi-messenger landscape and will allow multi-flavour neutrino alerts to be sent with unprecedented angular resolution.\\

\acknowledgments{
The authors acknowledge the financial support of the funding agencies:
Centre National de la Recherche Scientifique (CNRS), Commissariat \`a
l'\'ener\-gie atomique et aux \'energies alternatives (CEA),
Commission Europ\'eenne (FEDER fund and Marie Curie Program),
LabEx UnivEarthS (ANR-10-LABX-0023 and ANR-18-IDEX-0001),
R\'egion Alsace (contrat CPER), R\'egion Provence-Alpes-C\^ote d'Azur,
D\'e\-par\-tement du Var and Ville de La
Seyne-sur-Mer, France;
Bundesministerium f\"ur Bildung und Forschung
(BMBF), Germany; 
Istituto Nazionale di Fisica Nucleare (INFN), Italy;
Nederlandse organisatie voor Wetenschappelijk Onderzoek (NWO), the Netherlands;
Executive Unit for Financing Higher Education, Research, Development and Innovation (UEFISCDI), Romania;
MCIN for PID2021-124591NB-C41, -C42, -C43, funded by MCIN/AEI/10.13039/501100011033 and by “ERDF A way of making Europe”, for ASFAE/2022/014, ASFAE/2022 /023 and AST22\_6.2, with funding from the EU NextGenerationEU (PRTR-C17.I01), Generalitat Valenciana and Junta de Andaluc\'ia, and for CSIC-INFRA23013, Generalitat Valenciana for PROMETEO/2020/019, for CIDEGENT/2018/034, /2019/043, /2020/049, /2021/23 and for GRISOLIAP/2021/192 and EU for MSC/101025085, Spain;
Ministry of Higher Education, Scientific Research and Innovation, Morocco, and the Arab Fund for Economic and Social Development, Kuwait.
We also acknowledge the technical support of Ifremer, AIM and Foselev Marine
for the sea operation and the CC-IN2P3 for the computing facilities.
Based on observations collected at the European Organisation for Astronomical Research in the Southern Hemisphere under ESO programme(s) 095.D-0072(A). David Kaplan is supported by NSF grant AST-1816492.
}

\bibliographystyle{ieeetr}
\bibliography{references}

\begin{thebibliography}{10}

\bibitem{Aartsen:2016nxy}
M.~Aartsen {\em et~al.}, ``{The IceCube Neutrino Observatory: Instrumentation and Online Systems},'' {\em JINST}, vol.~12, no.~03, p.~P03012, 2017.

\bibitem{Collaboration:2011nsa}
M.~Ageron {\em et~al.}, ``{ANTARES: the first undersea neutrino telescope},'' {\em Nucl. Instrum. Meth. A}, vol.~656, pp.~11--38, 2011.

\bibitem{2020PAN....83..916A}
A.~V. {Avrorin}, A.~D. {Avrorin}, V.~M. {Aynutdinov}, R.~{Bannasch}, Z.~{Bardacova}, I.~A. {Belolaptikov}, V.~B. {Brudanin}, N.~M. {Budnev}, A.~R. {Gafarov}, K.~V. {Golubkov}, N.~S. {Gorshkov}, T.~I. {Gres}, R.~{Dvornicky}, G.~V. {Domogatsky}, A.~A. {Doroshenko}, Z.~A.~M. {Dzhilkibaev}, V.~Y. {Dik}, A.~N. {Dyachok}, E.~{Eckerova}, D.~N. {Zaborov}, R.~A. {Ivanov}, M.~S. {Katulin}, K.~G. {Kebkal}, O.~G. {Kebkal}, V.~A. {Kozhin}, M.~M. {Kolbin}, K.~V. {Konishev}, A.~V. {Korobchenko}, A.~P. {Koshechkin}, M.~V. {Kruglov}, M.~K. {Krjukov}, V.~F. {Kulepov}, M.~B. {Milenin}, R.~A. {Mirgazov}, V.~{Nazari}, D.~V. {Naumov}, A.~I. {Panfilov}, D.~P. {Petukhov}, E.~N. {Pliskovsky}, M.~I. {Rozanov}, V.~D. {Rushay}, E.~V. {Rjabov}, G.~B. {Safronov}, F.~{Simkovic}, A.~V. {Skurikhin}, A.~G. {Solovjev}, M.~N. {Sorokovikov}, I.~{Stekl}, O.~V. {Suvorova}, E.~O. {Sushenok}, V.~A. {Tabolenko}, B.~A. {Taraschansky}, L.~{Fajt}, S.~V. {Fialkovsky}, E.~V. {Khramov}, B.~A. {Shaibonov}, M.~D. {Shelepov}, and S.~A. {Yakovlyev},
  ``{Baikal-GVD Experiment},'' {\em Physics of Atomic Nuclei}, vol.~83, pp.~916--921, Nov. 2020.

\bibitem{Aartsen:2014gkd}
M.~Aartsen {\em et~al.}, ``{Observation of High-Energy Astrophysical Neutrinos in Three Years of IceCube Data},'' {\em Phys. Rev. Lett.}, vol.~113, p.~101101, 2014.

\bibitem{IceCube:2018dnn}
M.~Aartsen {\em et~al.}, ``{Multimessenger observations of a flaring blazar coincident with high-energy neutrino IceCube-170922A},'' {\em Science}, vol.~361, no.~6398, p.~eaat1378, 2018.

\bibitem{IceCube:2018cha}
M.~G. Aartsen {\em et~al.}, ``{Neutrino emission from the direction of the blazar TXS 0506+056 prior to the IceCube-170922A alert},'' {\em Science}, vol.~361, no.~6398, pp.~147--151, 2018.

\bibitem{Albert:2018kjg}
A.~Albert {\em et~al.}, ``{The Search for Neutrinos from TXS 0506+056 with the ANTARES Telescope},'' {\em Astrophys. J. Lett.}, vol.~863, no.~2, p.~L30, 2018.

\bibitem{Ageron:2011pe}
M.~Ageron {\em et~al.}, ``{The ANTARES Telescope Neutrino Alert System},'' {\em Astropart. Phys.}, vol.~35, pp.~530--536, 2012.

\bibitem{Adrian-Martinez:2015nin}
S.~Adrián-Martínez {\em et~al.}, ``{Optical and X-ray early follow-up of ANTARES neutrino alerts},'' {\em JCAP}, vol.~02, p.~062, 2016.

\bibitem{Aguilar:2011zz}
J.~Aguilar {\em et~al.}, ``{A fast algorithm for muon track reconstruction and its application to the ANTARES neutrino telescope},'' {\em Astropart. Phys.}, vol.~34, pp.~652--662, 2011.

\bibitem{AdrianMartinez:2012rp}
S.~Adrian-Martinez {\em et~al.}, ``{Search for Cosmic Neutrino Point Sources with Four Year Data of the ANTARES Telescope},'' {\em Astrophys. J.}, vol.~760, p.~53, 2012.

\bibitem{2012JInst...7T8002A}
S.~{Adri{\'a}n-Mart{\'\i}nez}, M.~{Ageron}, J.~A. {Aguilar}, I.~A. {Samarai}, A.~{Albert}, M.~{Andr{\'e}}, M.~{Anghinolfi}, G.~{Anton}, S.~{Anvar}, M.~{Ardid}, A.~C. {Assis Jesus}, T.~{Astraatmadja}, J.~J. {Aubert}, B.~{Baret}, S.~{Basa}, V.~{Bertin}, S.~{Biagi}, A.~{Bigi}, C.~{Bigongiari}, C.~{Bogazzi}, M.~{Bou-Cabo}, B.~{Bouhou}, M.~C. {Bouwhuis}, J.~{Brunner}, J.~{Busto}, F.~{Camarena}, A.~{Capone}, C.~{C{\^a}rloganu}, G.~{Carminati}, J.~{Carr}, S.~{Cecchini}, Z.~{Charif}, P.~{Charvis}, T.~{Chiarusi}, M.~{Circella}, R.~{Coniglione}, H.~{Costantini}, P.~{Coyle}, C.~{Curtil}, G.~{De Bonis}, M.~P. {Decowski}, I.~{Dekeyser}, A.~{Deschamps}, C.~{Distefano}, C.~{Donzaud}, D.~{Dornic}, Q.~{Dorosti}, D.~{Drouhin}, T.~{Eberl}, U.~{Emanuele}, A.~{Enzenh{\"o}fer}, J.~P. {Ernenwein}, S.~{Escoffier}, P.~{Fermani}, M.~{Ferri}, V.~{Flaminio}, F.~{Folger}, U.~{Fritsch}, J.~L. {Fuda}, S.~{Galat{\`a}}, P.~{Gay}, G.~{Giacomelli}, V.~{Giordano}, J.~P. {G{\'o}mez-Gonz{\'a}lez}, K.~{Graf}, G.~{Guillard}, G.~{Halladjian},
  G.~{Hallewell}, H.~{van Haren}, J.~{Hartman}, A.~J. {Heijboer}, Y.~{Hello}, J.~J. {Hern{\'a}ndez-Rey}, B.~{Herold}, J.~{H{\"o}{\ss}l}, C.~C. {Hsu}, M.~{de Jong}, M.~{Kadler}, O.~{Kalekin}, A.~{Kappes}, U.~{Katz}, O.~{Kavatsyuk}, P.~{Keller}, P.~{Kooijman}, C.~{Kopper}, A.~{Kouchner}, I.~{Kreykenbohm}, V.~{Kulikovskiy}, R.~{Lahmann}, P.~{Lamare}, G.~{Larosa}, D.~{Lattuada}, D.~{Lef{\`e}vre}, A.~{Le Van Suu}, G.~{Lim}, D.~{Lo Presti}, H.~{Loehner}, S.~{Loucatos}, S.~{Mangano}, M.~{Marcelin}, A.~{Margiotta}, J.~A. {Mart{\'\i}nez-Mora}, A.~{Meli}, T.~{Montaruli}, L.~{Moscoso}, H.~{Motz}, M.~{Neff}, E.~{Nezri}, V.~{Niess}, D.~{Palioselitis}, G.~E. {P{\u{a}}v{\u{a}}la{\c{s}}}, K.~{Payet}, P.~{Payre}, J.~{Petrovic}, P.~{Piattelli}, N.~{Picot-Clemente}, V.~{Popa}, T.~{Pradier}, E.~{Presani}, C.~{Racca}, D.~{Real}, C.~{Reed}, G.~{Riccobene}, C.~{Richardt}, R.~{Richter}, C.~{Rivi{\`e}re}, A.~{Robert}, K.~{Roensch}, A.~{Rostovtsev}, J.~{Ruiz-Rivas}, M.~{Rujoiu}, G.~V. {Russo}, F.~{Salesa}, D.~F.~E. {Samtleben},
  F.~{Sch{\"o}ck}, J.~P. {Schuller}, F.~{Sch{\"u}ssler}, T.~{Seitz}, R.~{Shanidze}, F.~{Simeone}, A.~{Spies}, M.~{Spurio}, J.~J.~M. {Steijger}, T.~{Stolarczyk}, A.~{S{\'a}nchez-Losa}, M.~{Taiuti}, C.~{Tamburini}, S.~{Toscano}, B.~{Vallage}, V.~{Van Elewyck}, G.~{Vannoni}, M.~{Vecchi}, P.~{Vernin}, S.~{Wagner}, G.~{Wijnker}, J.~{Wilms}, E.~{de Wolf}, H.~{Yepes}, D.~{Zaborov}, J.~D. {Zornoza}, and J.~{Z{\'u}{\~n}iga}, ``{The positioning system of the ANTARES Neutrino Telescope},'' {\em Journal of Instrumentation}, vol.~7, p.~T08002, Aug. 2012.

\bibitem{White:2011qf}
D.~J. White, E.~Daw, and V.~Dhillon, ``{A List of Galaxies for Gravitational Wave Searches},'' {\em Class. Quant. Grav.}, vol.~28, p.~085016, 2011.

\bibitem{GCN2000}
S.~Barthelmy, T.~L. Cline, R.~M. Butterworth, P.~Kippen, M.~S. Briggs, V.~Connaughton, and G.~N. Pendleton, ``Grb coordinates network (gcn): A status report,'' in {\em Gamma-ray Bursts, 5th Huntsville Symposium}, vol.~526 of {\em American Institute of Physics Conference Series}, pp.~731--735, 09 2000.

\bibitem{2011ivoa.spec.0711S}
R.~Seaman {\em et~al.}, ``{IVOA Recommendation: Sky Event Reporting Metadata Version 2.0},'' 10 2011.

\bibitem{Klotz_2009}
A.~Klotz, M.~Boer, J.~L. Atteia, and B.~Gendre, ``{Early optical observations of GRBs by the TAROT telescopes: period 2001-2008},'' {\em Astron. J.}, vol.~137, p.~4100, 2009.

\bibitem{ROTSE2003}
C.~W. Akerlof {\em et~al.}, ``{The ROTSE-III robotic telescope system},'' {\em Publ. Astron. Soc. Pac.}, vol.~115, pp.~132--140, 2003.

\bibitem{2003AAS...202.4702L}
V.~M. {Lipunov}, ``{Mobile Astronomical Systems of the Telescope-Robots (MASTER) near the Moscow},'' in {\em American Astronomical Society Meeting Abstracts \#202}, vol.~202 of {\em American Astronomical Society Meeting Abstracts}, p.~47.02, May 2003.

\bibitem{2005Ap.....48..389L}
V.~M. Lipunov {\em et~al.}, ``{MASTER: The Mobile Astronomical System of Telescope-Robots},'' {\em Astron. Nachr.}, vol.~325, pp.~580--582, 2004.

\bibitem{2010AdAst2010E..30L}
V.~Lipunov {\em et~al.}, ``{Master Robotic Net},'' {\em Adv. Astron.}, vol.~2010, p.~349171, 2010.

\bibitem{2012ExA....33..173K}
V.~Kornilov {\em et~al.}, ``{Robotic optical telescopes global network MASTER II. Equipment, structure, algorithms},'' {\em Exper. Astron.}, vol.~33, p.~173, 2012.

\bibitem{2020ApJ...896L..19L}
V.~M. Lipunov {\em et~al.}, ``{Optical Observations Reveal Strong Evidence for High Energy Neutrino Progenitor},'' {\em Astrophys. J. Lett.}, vol.~896, p.~L19, 06 2020.

\bibitem{Coward:2016jja}
D.~Coward {\em et~al.}, ``{The Zadko Telescope: Exploring the transient Universe},'' {\em Publ. Astron. Soc. Austral.}, vol.~34, p.~e005, 2017.

\bibitem{Wei:2016eox}
J.~Wei and B.~Cordier, ``{The Deep and Transient Universe in the SVOM Era: New Challenges and Opportunities - Scientific prospects of the SVOM mission},'' {\em arXiv e-prints}, 10 2016.

\bibitem{Gehrels:2004aa}
N.~Gehrels {\em et~al.}, ``{The Swift Gamma-Ray Burst Mission},'' {\em Astrophys. J.}, vol.~611, pp.~1005--1020, 2004.
\newblock [Erratum: Astrophys.J. 621, 558 (2005)].

\bibitem{Burrows:2008ts}
D.~N. Burrows {\em et~al.}, ``{The Swift X-Ray Telescope: Status and Performance},'' {\em Proc. SPIE Int. Soc. Opt. Eng.}, vol.~6686, p.~668607, 2007.

\bibitem{Evans_2013}
P.~A. Evans, J.~P. Osborne, A.~P. Beardmore, K.~L. Page, R.~Willingale, C.~J. Mountford, C.~Pagani, D.~N. Burrows, J.~A. Kennea, M.~Perri, G.~Tagliaferri, and N.~Gehrels, ``1sxps: A deep swift x-ray telescope point source catalog with light curves and spectra,'' {\em The Astrophysical Journal Supplement Series}, vol.~210, no.~1, p.~8, 2013.

\bibitem{61ccea70e38a49898265c91a1720cb49}
P.~A. Evans, J.~Osborne, J.~Kennea, M.~Smith, D.~Palmer, N.~Gehrels, J.~Gelbord, A.~Homeier, M.~Voge, N.~Strotjohann, D.~Cowen, S.~B{\"o}ser, M.~Kowalski, and A.~Stasik, ``Swift follow-up of icecube triggers, and implications for the advanced-ligo era,'' {\em Monthly Notices of the Royal Astronomical Society}, vol.~448, no.~3, pp.~2210--2223, 2015.

\bibitem{2003A&A...411L...1W}
C.~Winkler {\em et~al.}, ``{The INTEGRAL mission},'' {\em Astron. Astrophys.}, vol.~411, pp.~L1--L6, 2003.

\bibitem{2009IEEEP..97.1497L}
C.~J. Lonsdale {\em et~al.}, ``{The Murchison Widefield Array: Design Overview},'' {\em IEEE Proc.}, vol.~97, p.~1497, 2009.

\bibitem{2013PASA307T}
S.~J. Tingay {\em et~al.}, ``{The Murchison Widefield Array: the Square Kilometre Array Precursor at low radio frequencies},'' {\em Publ. Astron. Soc. Austral.}, vol.~30, p.~7, 2013.

\bibitem{Aharonian:2006pe}
F.~Aharonian {\em et~al.}, ``{Observations of the Crab Nebula with H.E.S.S},'' {\em Astron. Astrophys.}, vol.~457, pp.~899--915, 2006.

\bibitem{Schlegel:1997yv}
D.~J. Schlegel, D.~P. Finkbeiner, and M.~Davis, ``{Maps of dust IR emission for use in estimation of reddening and CMBR foregrounds},'' {\em Astrophys. J.}, vol.~500, p.~525, 1998.

\bibitem{Evans:2007na}
P.~Evans {\em et~al.}, ``{An online repository of Swift/XRT light curves of $\gamma$-ray bursts},'' {\em Astron. Astrophys.}, vol.~469, pp.~379--385, 2007.

\bibitem{2017APh....92...30A}
M.~G. {Aartsen}, M.~{Ackermann}, J.~{Adams}, J.~A. {Aguilar}, M.~{Ahlers}, M.~{Ahrens}, D.~{Altmann}, K.~{Andeen}, T.~{Anderson}, I.~{Ansseau}, G.~{Anton}, M.~{Archinger}, C.~{Arg{\"u}elles}, J.~{Auffenberg}, S.~{Axani}, X.~{Bai}, S.~W. {Barwick}, V.~{Baum}, R.~{Bay}, J.~J. {Beatty}, J.~{Becker Tjus}, K.~H. {Becker}, S.~{BenZvi}, D.~{Berley}, E.~{Bernardini}, A.~{Bernhard}, D.~Z. {Besson}, G.~{Binder}, D.~{Bindig}, M.~{Bissok}, E.~{Blaufuss}, S.~{Blot}, C.~{Bohm}, M.~{B{\"o}rner}, F.~{Bos}, D.~{Bose}, S.~{B{\"o}ser}, O.~{Botner}, J.~{Braun}, L.~{Brayeur}, H.~P. {Bretz}, S.~{Bron}, A.~{Burgman}, T.~{Carver}, M.~{Casier}, E.~{Cheung}, D.~{Chirkin}, A.~{Christov}, K.~{Clark}, L.~{Classen}, S.~{Coenders}, G.~H. {Collin}, J.~M. {Conrad}, D.~F. {Cowen}, R.~{Cross}, M.~{Day}, J.~P.~A.~M. {de Andr{\'e}}, C.~{De Clercq}, E.~{del Pino Rosendo}, H.~{Dembinski}, S.~{De Ridder}, P.~{Desiati}, K.~D. {de Vries}, G.~{de Wasseige}, M.~{de With}, T.~{DeYoung}, J.~C. {D{\'\i}az-V{\'e}lez}, V.~{di Lorenzo}, H.~{Dujmovic}, J.~P.
  {Dumm}, M.~{Dunkman}, B.~{Eberhardt}, T.~{Ehrhardt}, B.~{Eichmann}, P.~{Eller}, S.~{Euler}, P.~A. {Evenson}, S.~{Fahey}, A.~R. {Fazely}, J.~{Feintzeig}, J.~{Felde}, K.~{Filimonov}, C.~{Finley}, S.~{Flis}, C.~C. {F{\"o}sig}, A.~{Franckowiak}, E.~{Friedman}, T.~{Fuchs}, T.~K. {Gaisser}, J.~{Gallagher}, L.~{Gerhardt}, K.~{Ghorbani}, W.~{Giang}, L.~{Gladstone}, T.~{Glauch}, T.~{Gl{\"u}senkamp}, A.~{Goldschmidt}, J.~G. {Gonzalez}, D.~{Grant}, Z.~{Griffith}, C.~{Haack}, A.~{Hallgren}, F.~{Halzen}, E.~{Hansen}, T.~{Hansmann}, K.~{Hanson}, D.~{Hebecker}, D.~{Heereman}, K.~{Helbing}, R.~{Hellauer}, S.~{Hickford}, J.~{Hignight}, G.~C. {Hill}, K.~D. {Hoffman}, R.~{Hoffmann}, K.~{Hoshina}, F.~{Huang}, M.~{Huber}, K.~{Hultqvist}, S.~{In}, A.~{Ishihara}, E.~{Jacobi}, G.~S. {Japaridze}, M.~{Jeong}, K.~{Jero}, B.~J.~P. {Jones}, W.~{Kang}, A.~{Kappes}, T.~{Karg}, A.~{Karle}, U.~{Katz}, M.~{Kauer}, A.~{Keivani}, J.~L. {Kelley}, A.~{Kheirandish}, J.~{Kim}, M.~{Kim}, T.~{Kintscher}, J.~{Kiryluk}, T.~{Kittler}, S.~R. {Klein},
  G.~{Kohnen}, R.~{Koirala}, H.~{Kolanoski}, R.~{Konietz}, L.~{K{\"o}pke}, C.~{Kopper}, S.~{Kopper}, D.~J. {Koskinen}, M.~{Kowalski}, K.~{Krings}, M.~{Kroll}, G.~{Kr{\"u}ckl}, C.~{Kr{\"u}ger}, J.~{Kunnen}, S.~{Kunwar}, N.~{Kurahashi}, T.~{Kuwabara}, M.~{Labare}, J.~L. {Lanfranchi}, M.~J. {Larson}, F.~{Lauber}, D.~{Lennarz}, M.~{Lesiak-Bzdak}, M.~{Leuermann}, L.~{Lu}, J.~{L{\"u}nemann}, J.~{Madsen}, G.~{Maggi}, K.~B.~M. {Mahn}, S.~{Mancina}, M.~{Mandelartz}, R.~{Maruyama}, K.~{Mase}, R.~{Maunu}, F.~{McNally}, K.~{Meagher}, M.~{Medici}, M.~{Meier}, A.~{Meli}, T.~{Menne}, G.~{Merino}, T.~{Meures}, S.~{Miarecki}, T.~{Montaruli}, M.~{Moulai}, R.~{Nahnhauer}, U.~{Naumann}, G.~{Neer}, H.~{Niederhausen}, S.~C. {Nowicki}, D.~R. {Nygren}, A.~{Obertacke Pollmann}, A.~{Olivas}, A.~{O'Murchadha}, T.~{Palczewski}, H.~{Pandya}, D.~V. {Pankova}, P.~{Peiffer}, {\"O}.~{Penek}, J.~A. {Pepper}, C.~{P{\'e}rez de los Heros}, D.~{Pieloth}, E.~{Pinat}, P.~B. {Price}, G.~T. {Przybylski}, M.~{Quinnan}, C.~{Raab}, L.~{R{\"a}del},
  M.~{Rameez}, K.~{Rawlins}, R.~{Reimann}, B.~{Relethford}, M.~{Relich}, E.~{Resconi}, W.~{Rhode}, M.~{Richman}, B.~{Riedel}, S.~{Robertson}, M.~{Rongen}, C.~{Rott}, T.~{Ruhe}, D.~{Ryckbosch}, D.~{Rysewyk}, L.~{Sabbatini}, S.~E. {Sanchez Herrera}, A.~{Sandrock}, J.~{Sandroos}, S.~{Sarkar}, K.~{Satalecka}, P.~{Schlunder}, T.~{Schmidt}, S.~{Schoenen}, S.~{Sch{\"o}neberg}, L.~{Schumacher}, D.~{Seckel}, S.~{Seunarine}, D.~{Soldin}, M.~{Song}, G.~M. {Spiczak}, C.~{Spiering}, T.~{Stanev}, A.~{Stasik}, J.~{Stettner}, A.~{Steuer}, T.~{Stezelberger}, R.~G. {Stokstad}, A.~{St{\"o}{\ss}l}, R.~{Str{\"o}m}, N.~L. {Strotjohann}, G.~W. {Sullivan}, M.~{Sutherland}, H.~{Taavola}, I.~{Taboada}, J.~{Tatar}, F.~{Tenholt}, S.~{Ter-Antonyan}, A.~{Terliuk}, G.~{Te{\v{s}}i{\'c}}, S.~{Tilav}, P.~A. {Toale}, M.~N. {Tobin}, S.~{Toscano}, D.~{Tosi}, M.~{Tselengidou}, A.~{Turcati}, E.~{Unger}, M.~{Usner}, J.~{Vandenbroucke}, N.~{van Eijndhoven}, S.~{Vanheule}, M.~{van Rossem}, J.~{van Santen}, M.~{Vehring}, M.~{Voge}, E.~{Vogel},
  M.~{Vraeghe}, C.~{Walck}, A.~{Wallace}, M.~{Wallraff}, N.~{Wandkowsky}, C.~{Weaver}, M.~J. {Weiss}, C.~{Wendt}, S.~{Westerhoff}, B.~J. {Whelan}, S.~{Wickmann}, K.~{Wiebe}, C.~H. {Wiebusch}, L.~{Wille}, D.~R. {Williams}, L.~{Wills}, M.~{Wolf}, T.~R. {Wood}, E.~{Woolsey}, K.~{Woschnagg}, D.~L. {Xu}, X.~W. {Xu}, Y.~{Xu}, J.~P. {Yanez}, G.~{Yodh}, S.~{Yoshida}, and M.~{Zoll}, ``{The IceCube realtime alert system},'' {\em Astroparticle Physics}, vol.~92, pp.~30--41, June 2017.

\bibitem{Voges:1999ju}
W.~Voges {\em et~al.}, ``{The ROSAT all - sky survey bright source catalogue},'' {\em Astron. Astrophys.}, vol.~349, p.~389, 1999.

\bibitem{2015GCN.18231....1D}
D.~{Dornic}, S.~{Basa}, P.~A. {Evans}, J.~A. {Kennea}, J.~P. {Osborne}, and V.~{Lipunov}, ``{ANTARES neutrino detection and possible Swift X-ray counterpart.},'' {\em GRB Coordinates Network}, vol.~18231, p.~1, Jan. 2015.

\bibitem{2015ATel.7987....1D}
D.~{Dornic}, S.~{Basa}, P.~A. {Evans}, J.~A. {Kennea}, J.~P. {Osborne}, and V.~{Lipunov}, ``{ANTARES neutrino detection and possible Swift X-ray counterpart},'' {\em The Astronomer's Telegram}, vol.~7987, p.~1, Sept. 2015.

\bibitem{2015GCN.18240....1D}
D.~{Dornic}, V.~{Lipunov}, S.~{Basa}, P.~A. {Evans}, J.~A. {Kennea}, E.~{Gorbovskoy}, N.~{Tyurina}, D.~{Buckley}, and R.~{Rebolo}, ``{ANTARES neutrino Alert150901.32 alert and Swift XRT counterpart: MASTER optical observations and new possible candidate.},'' {\em GRB Coordinates Network}, vol.~18240, p.~1, Jan. 2015.

\bibitem{2015ATel.8000....1D}
D.~{Dornic}, V.~{Lipunov}, S.~{Basa}, P.~A. {Evans}, J.~A. {Kennea}, E.~{Gorbovskoy}, N.~{Tyurina}, D.~{Buckley}, and R.~{Rebolo}, ``{ANTARES high energy neutrino Alert150901.32 Error-box X-ray, B, V, R optical observations and Possible Candidate to Neutrino Source},'' {\em The Astronomer's Telegram}, vol.~8000, p.~1, Sept. 2015.

\bibitem{2015ATel.8027....1S}
S.~J. {Smartt}, K.~W. {Smith}, M.~{Huber}, K.~C. {Chambers}, H.~{Flewelling}, M.~{Willman}, N.~{Primak}, A.~{Schultz}, B.~{Gibson}, E.~{Magnier}, C.~{Waters}, J.~{Tonry}, R.~J. {Wainscoat}, D.~{Wright}, and D.~{Young}, ``{Pan-STARRS search for optical counterparts to the ANTARES neutrino detection},'' {\em The Astronomer's Telegram}, vol.~8027, p.~1, Sept. 2015.

\bibitem{2015ATel.8097....1K}
T.~{Kintscher} and A.~{Stasik}, ``{Search for Counterpart to ANTARES Neutrino Detection with IceCube},'' {\em The Astronomer's Telegram}, vol.~8097, p.~1, Sept. 2015.

\bibitem{2015GCN.18352....1B}
L.~{Blackburn}, M.~S. {Briggs}, E.~{Burns}, J.~{Camp}, N.~{Christensen}, V.~{Connaughton}, A.~{Goldstein}, P.~{Jenke}, T.~{Littenberg}, J.~{Racusin}, P.~{Shawhan}, L.~{Singer}, J.~{Veitch}, C.~{Wilson-Hodge}, and B.~{Zhang}, ``{ANTARES neutrino detection: Fermi GBM Observations.},'' {\em GRB Coordinates Network}, vol.~18352, p.~1, Jan. 2015.

\bibitem{2015ATel.7995....1F}
C.~{Ferrigno}, Y.~{Wang}, E.~{Kuulkers}, G.~{Belanger}, A.~{Bodaghee}, and J.~{Wilms}, ``{ANTARES neutrino detection: INTEGRAL upper limit on the hard X-ray counterpart},'' {\em The Astronomer's Telegram}, vol.~7995, p.~1, Sept. 2015.

\bibitem{2015ATel.8003....1N}
S.~{Nakahira}, S.~{Ueno}, H.~{Tomida}, M.~{Kimura}, M.~{Ishikawa}, Y.~E. {Nakagawa}, M.~{Sugizaki}, T.~{Mihara}, M.~{Serino}, M.~{Shidatsu}, J.~{Sugimoto}, T.~{Takagi}, M.~{Matsuoka}, N.~{Kawai}, M.~{Arimoto}, T.~{Yoshii}, Y.~{Tachibana}, Y.~{Ono}, T.~{Fujiwara}, A.~{Yoshida}, T.~{Sakamoto}, Y.~{Kawakubo}, H.~{Ohtsuki}, H.~{Tsunemi}, R.~{Imatani}, H.~{Negoro}, M.~{Nakajima}, K.~{Tanaka}, T.~{Masumitsu}, Y.~{Ueda}, T.~{Kawamuro}, T.~{Hori}, Y.~{Tsuboi}, S.~{Kanetou}, M.~{Yamauchi}, D.~{Itoh}, K.~{Yamaoka}, and M.~{Morii}, ``{MAXI/GSC upper limit for an X-ray counterpart of the ANTARES neutrino event detected on September 1.},'' {\em The Astronomer's Telegram}, vol.~8003, p.~1, Sept. 2015.

\bibitem{2017ICRC...35..653S}
F.~Sch\"ussler {\em et~al.}, ``{H.E.S.S. observations following multi-messenger alerts in real-time},'' in {\em 35th International Cosmic Ray Conference (ICRC2017)}, vol.~ICRC2017, p.~653, 2018.

\bibitem{2015ATel.8203....1M}
R.~{Mirzoyan}, ``{MAGIC observation of the ANTARES-detected neutrino sky region},'' {\em The Astronomer's Telegram}, vol.~8203, p.~1, Oct. 2015.

\bibitem{Hodapp2010}
K.~Hodapp, R.~Chini, B.~Reipurth, M.~Murphy, R.~Lemke, R.~Watermann, S.~Jacobson, K.~Bischoff, T.~Chonis, D.~Dement, R.~Terrien, and K.~Bott, ``Commissioning of the infrared imaging survey (iris) system,'' {\em Proceedings of SPIE - The International Society for Optical Engineering}, vol.~7735, 07 2010.

\bibitem{2015ATel.8006....1T}
J.~{Takahashi}, S.~{Narusawa}, S.~{Sai}, T.~{Hashimoto}, and {Nayuta Team}, ``{ANTARES neutrino detection: Nishi-Harima NIR photometry},'' {\em The Astronomer's Telegram}, vol.~8006, p.~1, Sept. 2015.

\bibitem{1992IAUS.149.225K}
R.~L. {Kurucz}, ``{Model Atmospheres for Population Synthesis},'' in {\em The Stellar Populations of Galaxies} (B.~{Barbuy} and A.~{Renzini}, eds.), vol.~149 of {\em IAU Symposium}, p.~225, Jan. 1992.

\bibitem{2015ATel.7994....1D}
A.~{de Ugarte Postigo}, H.~{Korhonen}, M.~I. {Andersen}, J.~P.~U. {Fynbo}, S.~{Schulze}, Z.~{Cano}, D.~{Xu}, N.~R. {Tanvir}, D.~{Watson}, D.~{Malesani}, J.~{Hjorth}, and A.~A. {Djupvik}, ``{ANTARES neutrino detection: Optical/NIR spectroscopy of the Swift/XRT counterpart candidate from NOT},'' {\em The Astronomer's Telegram}, vol.~7994, p.~1, Sept. 2015.

\bibitem{2015ATel.7993....1D}
S.~{Dichiara}, C.~{Koen}, T.~{Koen}, M.~{Kotze}, D.~{Milisavljevic}, R.~{Margutti}, and C.~{Guidorzi}, ``{ANTARES neutrino detection: optical spectroscopy of X-ray counterpart candidate with SALT},'' {\em The Astronomer's Telegram}, vol.~7993, p.~1, Sept. 2015.

\bibitem{2015ATel.7998....1C}
A.~J. {Castro-Tirado}, D.~{Galad{\'\i}-Enr{\'\i}quez}, F.~{Hoyos}, A.~{Guijarro}, R.~{S{\'a}nchez-Ram{\'\i}rez}, M.~{Fern{\'a}ndez}, J.~C. {Tello}, S.~{Jeong}, and J.~{Ma{\'\i}z-Apell{\'a}niz}, ``{ANTARES neutrino detection: CAHA photometry \& spectroscopy of the Swift source},'' {\em The Astronomer's Telegram}, vol.~7998, p.~1, Sept. 2015.

\bibitem{2015ATel.7996....1T}
B.~E. {Tucker}, K.~{Freeman}, F.~{Yuan}, C.~{Wolf}, B.~{Schmidt}, D.~{Bayliss}, and C.~{Onken}, ``{WiFeS and Kepler K2 Observations of the stellar x-ray source within the ANTARES neutrino detection region},'' {\em The Astronomer's Telegram}, vol.~7996, p.~1, Sept. 2015.

\bibitem{2015ATel.7999....1H}
G.~{Hallinan} and K.~{Kunal Mooley}, ``{Jansky VLA observation of the ANTARES neutrino detection region},'' {\em The Astronomer's Telegram}, vol.~7999, p.~1, Sept. 2015.

\bibitem{2015ATel.8034....1T}
A.~{Tetarenko}, G.~{Sivakoff}, A.~{Bahramian}, C.~O. {Heinke}, G.~{Hallinan}, J.~{Miller-Jones}, A.~{Mioduszewski}, and K.~{Mooley}, ``{ANTARES neutrino detection: A preliminary VLA catalogue of radio source components and their variability levels in the field},'' {\em The Astronomer's Telegram}, vol.~8034, p.~1, Sept. 2015.

\bibitem{2005PhRvL..95q1101A}
S.~{Ando}, J.~F. {Beacom}, and H.~{Y{\"u}ksel}, ``{Detection of Neutrinos from Supernovae in Nearby Galaxies},'' {\em prl}, vol.~95, p.~171101, Oct. 2005.

\bibitem{Massaro:2008ye}
E.~Massaro, P.~Giommi, C.~Leto, P.~Marchegiani, A.~Maselli, M.~Perri, S.~Piranomonte, and S.~Sclavi, ``{Roma-BZCAT: A multifrequency catalogue of Blazars},'' {\em Astron. Astrophys.}, vol.~495, p.~691, 2009.

\bibitem{Massaro:2015nia}
E.~Massaro, A.~Maselli, C.~Leto, P.~Marchegiani, M.~Perri, P.~Giommi, and S.~Piranomonte, ``{The 5th edition of the Roma-BZCAT. A short presentation},'' {\em Astrophys. Space Sci.}, vol.~357, no.~1, p.~75, 2015.

\bibitem{2021A&A...649A...1G}
{Gaia Collaboration}, A.~G.~A. {Brown}, A.~{Vallenari}, T.~{Prusti}, J.~H.~J. {de Bruijne}, C.~{Babusiaux}, M.~{Biermann}, O.~L. {Creevey}, D.~W. {Evans}, L.~{Eyer}, A.~{Hutton}, F.~{Jansen}, C.~{Jordi}, S.~A. {Klioner}, U.~{Lammers}, L.~{Lindegren}, X.~{Luri}, F.~{Mignard}, C.~{Panem}, D.~{Pourbaix}, S.~{Randich}, P.~{Sartoretti}, C.~{Soubiran}, N.~A. {Walton}, F.~{Arenou}, C.~A.~L. {Bailer-Jones}, U.~{Bastian}, M.~{Cropper}, R.~{Drimmel}, D.~{Katz}, M.~G. {Lattanzi}, F.~{van Leeuwen}, J.~{Bakker}, C.~{Cacciari}, J.~{Casta{\~n}eda}, F.~{De Angeli}, C.~{Ducourant}, C.~{Fabricius}, M.~{Fouesneau}, Y.~{Fr{\'e}mat}, R.~{Guerra}, A.~{Guerrier}, J.~{Guiraud}, A.~{Jean-Antoine Piccolo}, E.~{Masana}, R.~{Messineo}, N.~{Mowlavi}, C.~{Nicolas}, K.~{Nienartowicz}, F.~{Pailler}, P.~{Panuzzo}, F.~{Riclet}, W.~{Roux}, G.~M. {Seabroke}, R.~{Sordo}, P.~{Tanga}, F.~{Th{\'e}venin}, G.~{Gracia-Abril}, J.~{Portell}, D.~{Teyssier}, M.~{Altmann}, R.~{Andrae}, I.~{Bellas-Velidis}, K.~{Benson}, J.~{Berthier}, R.~{Blomme},
  E.~{Brugaletta}, P.~W. {Burgess}, G.~{Busso}, B.~{Carry}, A.~{Cellino}, N.~{Cheek}, G.~{Clementini}, Y.~{Damerdji}, M.~{Davidson}, L.~{Delchambre}, A.~{Dell'Oro}, J.~{Fern{\'a}ndez-Hern{\'a}ndez}, L.~{Galluccio}, P.~{Garc{\'\i}a-Lario}, M.~{Garcia-Reinaldos}, J.~{Gonz{\'a}lez-N{\'u}{\~n}ez}, E.~{Gosset}, R.~{Haigron}, J.~L. {Halbwachs}, N.~C. {Hambly}, D.~L. {Harrison}, D.~{Hatzidimitriou}, U.~{Heiter}, J.~{Hern{\'a}ndez}, D.~{Hestroffer}, S.~T. {Hodgkin}, B.~{Holl}, K.~{Jan{\ss}en}, G.~{Jevardat de Fombelle}, S.~{Jordan}, A.~{Krone-Martins}, A.~C. {Lanzafame}, W.~{L{\"o}ffler}, A.~{Lorca}, M.~{Manteiga}, O.~{Marchal}, P.~M. {Marrese}, A.~{Moitinho}, A.~{Mora}, K.~{Muinonen}, P.~{Osborne}, E.~{Pancino}, T.~{Pauwels}, J.~M. {Petit}, A.~{Recio-Blanco}, P.~J. {Richards}, M.~{Riello}, L.~{Rimoldini}, A.~C. {Robin}, T.~{Roegiers}, J.~{Rybizki}, L.~M. {Sarro}, C.~{Siopis}, M.~{Smith}, A.~{Sozzetti}, A.~{Ulla}, E.~{Utrilla}, M.~{van Leeuwen}, W.~{van Reeven}, U.~{Abbas}, A.~{Abreu Aramburu}, S.~{Accart},
  C.~{Aerts}, J.~J. {Aguado}, M.~{Ajaj}, G.~{Altavilla}, M.~A. {{\'A}lvarez}, J.~{{\'A}lvarez Cid-Fuentes}, J.~{Alves}, R.~I. {Anderson}, E.~{Anglada Varela}, T.~{Antoja}, M.~{Audard}, D.~{Baines}, S.~G. {Baker}, L.~{Balaguer-N{\'u}{\~n}ez}, E.~{Balbinot}, Z.~{Balog}, C.~{Barache}, D.~{Barbato}, M.~{Barros}, M.~A. {Barstow}, S.~{Bartolom{\'e}}, J.~L. {Bassilana}, N.~{Bauchet}, A.~{Baudesson-Stella}, U.~{Becciani}, M.~{Bellazzini}, M.~{Bernet}, S.~{Bertone}, L.~{Bianchi}, S.~{Blanco-Cuaresma}, T.~{Boch}, A.~{Bombrun}, D.~{Bossini}, S.~{Bouquillon}, A.~{Bragaglia}, L.~{Bramante}, E.~{Breedt}, A.~{Bressan}, N.~{Brouillet}, B.~{Bucciarelli}, A.~{Burlacu}, D.~{Busonero}, A.~G. {Butkevich}, R.~{Buzzi}, E.~{Caffau}, R.~{Cancelliere}, H.~{C{\'a}novas}, T.~{Cantat-Gaudin}, R.~{Carballo}, T.~{Carlucci}, M.~I. {Carnerero}, J.~M. {Carrasco}, L.~{Casamiquela}, M.~{Castellani}, A.~{Castro-Ginard}, P.~{Castro Sampol}, L.~{Chaoul}, P.~{Charlot}, L.~{Chemin}, A.~{Chiavassa}, M.~R.~L. {Cioni}, G.~{Comoretto}, W.~J. {Cooper},
  T.~{Cornez}, S.~{Cowell}, F.~{Crifo}, M.~{Crosta}, C.~{Crowley}, C.~{Dafonte}, A.~{Dapergolas}, M.~{David}, P.~{David}, P.~{de Laverny}, F.~{De Luise}, R.~{De March}, J.~{De Ridder}, R.~{de Souza}, P.~{de Teodoro}, A.~{de Torres}, E.~F. {del Peloso}, E.~{del Pozo}, M.~{Delbo}, A.~{Delgado}, H.~E. {Delgado}, J.~B. {Delisle}, P.~{Di Matteo}, S.~{Diakite}, C.~{Diener}, E.~{Distefano}, C.~{Dolding}, D.~{Eappachen}, B.~{Edvardsson}, H.~{Enke}, P.~{Esquej}, C.~{Fabre}, M.~{Fabrizio}, S.~{Faigler}, G.~{Fedorets}, P.~{Fernique}, A.~{Fienga}, F.~{Figueras}, C.~{Fouron}, F.~{Fragkoudi}, E.~{Fraile}, F.~{Franke}, M.~{Gai}, D.~{Garabato}, A.~{Garcia-Gutierrez}, M.~{Garc{\'\i}a-Torres}, A.~{Garofalo}, P.~{Gavras}, E.~{Gerlach}, R.~{Geyer}, P.~{Giacobbe}, G.~{Gilmore}, S.~{Girona}, G.~{Giuffrida}, R.~{Gomel}, A.~{Gomez}, I.~{Gonzalez-Santamaria}, J.~J. {Gonz{\'a}lez-Vidal}, M.~{Granvik}, R.~{Guti{\'e}rrez-S{\'a}nchez}, L.~P. {Guy}, M.~{Hauser}, M.~{Haywood}, A.~{Helmi}, S.~L. {Hidalgo}, T.~{Hilger}, N.~{H{\l}adczuk},
  D.~{Hobbs}, G.~{Holland}, H.~E. {Huckle}, G.~{Jasniewicz}, P.~G. {Jonker}, J.~{Juaristi Campillo}, F.~{Julbe}, L.~{Karbevska}, P.~{Kervella}, S.~{Khanna}, A.~{Kochoska}, M.~{Kontizas}, G.~{Kordopatis}, A.~J. {Korn}, Z.~{Kostrzewa-Rutkowska}, K.~{Kruszy{\'n}ska}, S.~{Lambert}, A.~F. {Lanza}, Y.~{Lasne}, J.~F. {Le Campion}, Y.~{Le Fustec}, Y.~{Lebreton}, T.~{Lebzelter}, S.~{Leccia}, N.~{Leclerc}, I.~{Lecoeur-Taibi}, S.~{Liao}, E.~{Licata}, E.~P. {Lindstr{\o}m}, T.~A. {Lister}, E.~{Livanou}, A.~{Lobel}, P.~{Madrero Pardo}, S.~{Managau}, R.~G. {Mann}, J.~M. {Marchant}, M.~{Marconi}, M.~M.~S. {Marcos Santos}, S.~{Marinoni}, F.~{Marocco}, D.~J. {Marshall}, L.~{Martin Polo}, J.~M. {Mart{\'\i}n-Fleitas}, A.~{Masip}, D.~{Massari}, A.~{Mastrobuono-Battisti}, T.~{Mazeh}, P.~J. {McMillan}, S.~{Messina}, D.~{Michalik}, N.~R. {Millar}, A.~{Mints}, D.~{Molina}, R.~{Molinaro}, L.~{Moln{\'a}r}, P.~{Montegriffo}, R.~{Mor}, R.~{Morbidelli}, T.~{Morel}, D.~{Morris}, A.~F. {Mulone}, D.~{Munoz}, T.~{Muraveva}, C.~P. {Murphy},
  I.~{Musella}, L.~{Noval}, C.~{Ord{\'e}novic}, G.~{Orr{\`u}}, J.~{Osinde}, C.~{Pagani}, I.~{Pagano}, L.~{Palaversa}, P.~A. {Palicio}, A.~{Panahi}, M.~{Pawlak}, X.~{Pe{\~n}alosa Esteller}, A.~{Penttil{\"a}}, A.~M. {Piersimoni}, F.~X. {Pineau}, E.~{Plachy}, G.~{Plum}, E.~{Poggio}, E.~{Poretti}, E.~{Poujoulet}, A.~{Pr{\v{s}}a}, L.~{Pulone}, E.~{Racero}, S.~{Ragaini}, M.~{Rainer}, C.~M. {Raiteri}, N.~{Rambaux}, P.~{Ramos}, M.~{Ramos-Lerate}, P.~{Re Fiorentin}, S.~{Regibo}, C.~{Reyl{\'e}}, V.~{Ripepi}, A.~{Riva}, G.~{Rixon}, N.~{Robichon}, C.~{Robin}, M.~{Roelens}, L.~{Rohrbasser}, M.~{Romero-G{\'o}mez}, N.~{Rowell}, F.~{Royer}, K.~A. {Rybicki}, G.~{Sadowski}, A.~{Sagrist{\`a} Sell{\'e}s}, J.~{Sahlmann}, J.~{Salgado}, E.~{Salguero}, N.~{Samaras}, V.~{Sanchez Gimenez}, N.~{Sanna}, R.~{Santove{\~n}a}, M.~{Sarasso}, M.~{Schultheis}, E.~{Sciacca}, M.~{Segol}, J.~C. {Segovia}, D.~{S{\'e}gransan}, D.~{Semeux}, S.~{Shahaf}, H.~I. {Siddiqui}, A.~{Siebert}, L.~{Siltala}, E.~{Slezak}, R.~L. {Smart}, E.~{Solano},
  F.~{Solitro}, D.~{Souami}, J.~{Souchay}, A.~{Spagna}, F.~{Spoto}, I.~A. {Steele}, H.~{Steidelm{\"u}ller}, C.~A. {Stephenson}, M.~{S{\"u}veges}, L.~{Szabados}, E.~{Szegedi-Elek}, F.~{Taris}, G.~{Tauran}, M.~B. {Taylor}, R.~{Teixeira}, W.~{Thuillot}, N.~{Tonello}, F.~{Torra}, J.~{Torra}, C.~{Turon}, N.~{Unger}, M.~{Vaillant}, E.~{van Dillen}, O.~{Vanel}, A.~{Vecchiato}, Y.~{Viala}, D.~{Vicente}, S.~{Voutsinas}, M.~{Weiler}, T.~{Wevers}, {\L}.~{Wyrzykowski}, A.~{Yoldas}, P.~{Yvard}, H.~{Zhao}, J.~{Zorec}, S.~{Zucker}, C.~{Zurbach}, and T.~{Zwitter}, ``{Gaia Early Data Release 3. Summary of the contents and survey properties},'' {\em aap}, vol.~649, p.~A1, May 2021.

\bibitem{Croft:2016lhf}
S.~Croft {\em et~al.}, ``{Murchison Widefield Array Limits on Radio Emission from ANTARES Neutrino Events},'' {\em Astrophys. J. Lett.}, vol.~820, no.~2, p.~L24, 2016.

\bibitem{2019PASA...36...46H}
P.~J. Hancock and others., ``{A VOEvent-based automatic trigger system for the Murchison Widefield Array},'' {\em Publ. Astron. Soc. Austral.}, vol.~36, p.~46, 11 2019.

\bibitem{2019A&C....27...23H}
P.~J. {Hancock}, N.~{Hurley-Walker}, and T.~E. {White}, ``{ROBBIE: A batch processing work-flow for the detection of radio transients and variables},'' {\em Astronomy and Computing}, vol.~27, p.~23, Apr. 2019.

\bibitem{2018ApJ...867L..12S}
M.~Sokolowski {\em et~al.}, ``{No low-frequency emission from extremely bright Fast Radio Bursts},'' {\em Astrophys. J. Lett.}, vol.~867, no.~1, p.~L12, 2018.

\bibitem{2017MNRAS.466.1944M}
T.~{Murphy}, D.~L. {Kaplan}, S.~{Croft}, C.~{Lynch}, J.~R. {Callingham}, K.~{Bannister}, M.~E. {Bell}, N.~{Hurley-Walker}, P.~{Hancock}, J.~{Line}, A.~{Rowlinson}, E.~{Lenc}, H.~T. {Intema}, P.~{Jagannathan}, R.~D. {Ekers}, S.~{Tingay}, F.~{Yuan}, C.~{Wolf}, C.~A. {Onken}, K.~S. {Dwarakanath}, B.~Q. {For}, B.~M. {Gaensler}, L.~{Hindson}, M.~{Johnston-Hollitt}, A.~D. {Kapi{\'n}ska}, B.~{McKinley}, J.~{Morgan}, A.~R. {Offringa}, P.~{Procopio}, L.~{Staveley-Smith}, R.~{Wayth}, C.~{Wu}, and Q.~{Zheng}, ``{A search for long-time-scale, low-frequency radio transients},'' {\em {Mon. Not. Roy. Astron. Soc.}}, vol.~466, pp.~1944--1953, Apr. 2017.

\bibitem{2017MNRAS.464.1146H}
N.~{Hurley-Walker}, J.~R. {Callingham}, P.~J. {Hancock}, T.~M.~O. {Franzen}, L.~{Hindson}, A.~D. {Kapi{\'n}ska}, J.~{Morgan}, A.~R. {Offringa}, R.~B. {Wayth}, C.~{Wu}, Q.~{Zheng}, T.~{Murphy}, M.~E. {Bell}, K.~S. {Dwarakanath}, B.~{For}, B.~M. {Gaensler}, M.~{Johnston-Hollitt}, E.~{Lenc}, P.~{Procopio}, L.~{Staveley-Smith}, R.~{Ekers}, J.~D. {Bowman}, F.~{Briggs}, R.~J. {Cappallo}, A.~A. {Deshpande}, L.~{Greenhill}, B.~J. {Hazelton}, D.~L. {Kaplan}, C.~J. {Lonsdale}, S.~R. {McWhirter}, D.~A. {Mitchell}, M.~F. {Morales}, E.~{Morgan}, D.~{Oberoi}, S.~M. {Ord}, T.~{Prabu}, N.~U. {Shankar}, K.~S. {Srivani}, R.~{Subrahmanyan}, S.~J. {Tingay}, R.~L. {Webster}, A.~{Williams}, and C.~L. {Williams}, ``{GaLactic and Extragalactic All-sky Murchison Widefield Array (GLEAM) survey - I. A low-frequency extragalactic catalogue},'' {\em {Mon. Not. Roy. Astron. Soc.}}, vol.~464, pp.~1146--1167, Jan. 2017.

\bibitem{2017AIPC.1792f0006S}
F.~Sch\"ussler {\em et~al.}, ``{The H.E.S.S. multi-messenger program: Searches for TeV gamma-ray emission associated with high-energy neutrinos},'' in {\em 6th International Symposium on High Energy Gamma-Ray Astronomy}, vol.~1792 of {\em American Institute of Physics Conference Series}, p.~060006, 01 2017.

\bibitem{IceCube:2016ipa}
M.~G. Aartsen {\em et~al.}, ``{An All-Sky Search for Three Flavors of Neutrinos from Gamma-Ray Bursts with the IceCube Neutrino Observatory},'' {\em Astrophys. J.}, vol.~824, no.~2, p.~115, 2016.

\bibitem{ANTARES:2020dpd}
A.~Albert {\em et~al.}, ``{ANTARES upper limits on the multi-TeV neutrino emission from the GRBs detected by IACTs},'' {\em JCAP}, vol.~03, p.~092, 2021.

\bibitem{Senno:2015tsn}
N.~Senno, K.~Murase, and P.~Meszaros, ``{Choked Jets and Low-Luminosity Gamma-Ray Bursts as Hidden Neutrino Sources},'' {\em Phys. Rev. D}, vol.~93, no.~8, p.~083003, 2016.

\bibitem{Murase:2013ffa}
K.~Murase and K.~Ioka, ``{TeV\textendash{}PeV Neutrinos from Low-Power Gamma-Ray Burst Jets inside Stars},'' {\em Phys. Rev. Lett.}, vol.~111, no.~12, p.~121102, 2013.

\bibitem{KM3Net:2016zxf}
S.~Adrian-Martinez {\em et~al.}, ``{Letter of intent for KM3NeT 2.0},'' {\em J. Phys. G}, vol.~43, no.~8, p.~084001, 2016.

\end{thebibliography}

\end{document}